\documentclass[letterpaper]{article} 
\usepackage{aaai25}  
\usepackage{times}  
\usepackage{helvet}  
\usepackage{courier}  
\usepackage[hyphens]{url}  
\usepackage{graphicx} 
\usepackage{natbib}  
\usepackage{caption} 
\usepackage{algorithm}
\usepackage{algorithmic}

\usepackage{newfloat}
\usepackage{listings}
\DeclareCaptionStyle{ruled}{labelfont=normalfont,labelsep=colon,strut=off} 
\floatstyle{ruled}
\newfloat{listing}{tb}{lst}{}
\floatname{listing}{Listing}
 \nocopyright 
\usepackage{makecell}

\usepackage{xcolor}
\newcommand{\answerYes}[1]{\textcolor{blue}{#1}} 
\newcommand{\answerNo}[1]{\textcolor{teal}{#1}} 
\newcommand{\answerNA}[1]{\textcolor{gray}{#1}}

\newcommand{\newtext}[1]{\textcolor{black}{#1}} 

\newcommand{\deltext}[1]{\textcolor{red}{#1}} 
\renewcommand{\deltext}[1]{}
\title{Using Salient Object Detection to Identify Manipulative Cookie Banners that Circumvent GDPR}
\author{
    Riley Grossman,
    Michael Smith,
    Cristian Borcea,
    Yi Chen,
}
\affiliations{
    New Jersey Institute of Technology, Newark, NJ\\

    rag24@njit.edu, mes6@njit.edu, borcea@njit.edu, yi.chen@njit.edu
}

\usepackage{bibentry}

\newcommand{\yc}[1]{\textcolor{red}{#1}}
\renewcommand{\yc}[1]{}

\begin{document}

\maketitle
\begin{abstract}
The main goal of this paper is to study how often cookie banners that comply with the General Data Protection Regulation (GDPR) contain aesthetic manipulation,  a design tactic to draw users' attention to the button that permits personal data sharing.
As a byproduct of this goal, we also evaluate how frequently the banners comply with GDPR and the recommendations of national data protection authorities regarding banner designs. We visited 2,579 websites and identified the type of cookie banner implemented. Although 45\% of the relevant websites have fully compliant banners, we found aesthetic manipulation on \newtext{38\%} of the compliant banners.
\yc{what do you mean by "relevant"}
Unlike prior studies of aesthetic manipulation, we use a computer vision model for salient object detection to measure how salient (i.e., attention-drawing) each banner element is. This enables the discovery of new types of aesthetic manipulation (e.g., button placement), and leads us to conclude that aesthetic manipulation is more common than previously reported \newtext{(38.0\% vs 27.0\% of banners)}.
To study the effects of user and/or website location on cookie banner design, we include websites within the European Union (EU), where privacy regulation enforcement is more stringent, and websites outside the EU. We visited websites from IP addresses in the EU and from IP addresses in the United States (US). We find that 13.9\% of EU websites  change their banner design when the user is from the US, and EU websites are roughly \newtext{48.3\%} more likely to use aesthetic manipulation than non-EU websites, \newtext{highlighting their innovative responses to privacy regulation.} 

\end{abstract}

%
 \begin{links}
     \link{Data and Code}{https://github.com/rag24/aesthetic-manipulation-SOD}
 \end{links}

\section{Introduction}

\begin{figure}[t!]
\centering
\includegraphics[width=0.9\linewidth]{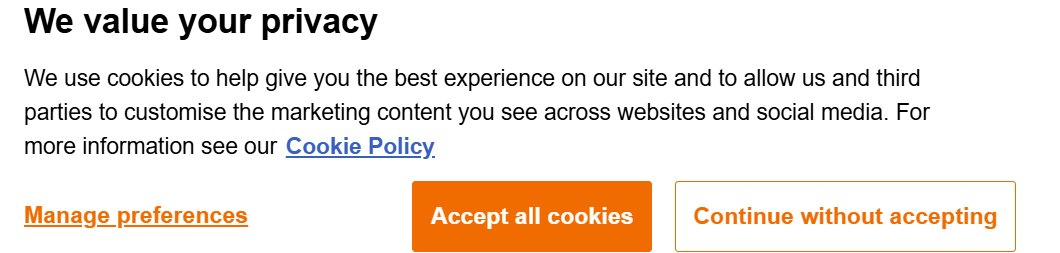}
\caption{Cookie Banner} 
\label{fig:notice}
\end{figure}

Privacy regulations to protect citizens' personal data have changed the web browsing experience. Following the enforcement of comprehensive privacy regulations such as the GDPR of the European Union (EU) in 2018, web users started to see more notices such as the one shown in Figure~\ref{fig:notice}. GDPR requires all companies that collect and/or process personal data to be transparent with the data subjects. Additionally, outside of specific circumstances (e.g., legal obligations), GDPR requires the companies to obtain consent from the data subject. These notices, known as cookie banners, help websites meet these requirements. They are a critical component of privacy on the Web. In theory, they return data ownership to the web user and allow each individual to decide how and when they want their personal data to be shared and used. 

\yc{I would think cookie banner is neutrla}

However, websites collect personal data to increase the revenue they receive from advertisers who want to target specific users with ads based on users' personal data. This puts websites' incentives in conflict with cookie banners that enable users to tell the website not to collect their data. This conflict explains why early studies found that cookie banners do not aid user privacy as intended~\cite{utz,degeling2019}. These early results were influenced by the fact that GDPR did not set cookie banner requirements and enforcement was lacking. Slowly, rulings from the European Court of Justice~\cite{planet49} and fines for non-compliant cookie banner designs~\cite{cnilfines} appeared. This resulted in more cookie banners that were privacy-friendly (e.g., required to place the ``reject'' button next to the ``accept'' button instead of hiding it in the settings). 

Nevertheless, websites still want to collect as much personal data as possible without being fined for non-compliance. 
This has led to the proliferation of dark patterns in cookie banner designs that nudge users toward sharing personal data. This is a problem because these dark patterns, such as aesthetic manipulation, are a way for companies to avoid fines by complying with the GDPR while still harming user privacy on the Web. 
The literature~\cite{gray} has defined aesthetic manipulation as banner designs that make the accept button more visually appealing (i.e., more salient) than other buttons. In related work, aesthetic manipulation is often operationalized as a highlighted accept button, as shown in Figure~\ref{fig:notice}. However, we believe that previous work has underreported the prevalence of aesthetic manipulation, as it did not consider more sophisticated design choices such as button placement. Furthermore, the previous work did not study the effect of user and website location on aesthetic manipulation.

\yc{any other literature beside \cite{gray}?}

The aim of this paper is to reliably study aesthetic manipulation in cookie banner design. Toward this goal, we identify a new method for measuring the prevalence of aesthetic manipulation in cookie banners on the Web. Specifically, we borrow from the computer vision field of salient object detection (SOD), and use a model called \textit{DeepRare}~\cite{dr21} to directly determine how salient each button on cookie banners is. Using DeepRare to directly calculate button salience, instead of focusing on a single design tactic (i.e., button highlighting), allowed us to capture all forms of aesthetic manipulation. This led to our conclusion that aesthetic manipulation occurs on \newtext{38\%} of the \newtext{GDPR-}compliant cookie banners. Methods from prior works would have only identified aesthetic manipulation in \newtext{27\%} of the same banners, \newtext{showing that prior work underestimated the prevalence of aesthetic manipulation}.
We then collect cookie banner and button characteristics to determine which factors (in addition to button highlighting) can affect salience. We find that displaying a button as a link and button placement \newtext{and size are all} significantly correlated with button salience and should be considered forms of aesthetic manipulation.


As part of our main goal to more thoroughly investigate aesthetic manipulation, we needed to first determine how many cookie banner designs were compliant with GDPR and the related guidelines set by EU data protection authorities~\cite{noyb,cnilrecs}. There is a lack of recent studies that measure the effects of user and website location on cookie banner design. Although one study has done this~\cite{warberg}, the data was collected in 2018-2019 (before any fines were given for non-compliant cookie banners) and the categories of banner designs were broad. To address this gap, we collect a large dataset of cookie banner images, which we make publicly available, by visiting 2,579 EU and non-EU websites (of varying popularity) from EU and US IP addresses. We then classify the banners into 17 different types of cookie banner designs, which is more detailed than prior work and allows us to report on the prevalence of new designs (e.g., corner reject button design). This dataset allows us to evaluate how the website and user locations affect cookie banner design. Analysis on this large dataset demonstrates that 13.9\% of EU websites show a more compliant banner to EU users than US users\newtext{, despite GDPR applying in both cases}. 

\newtext{To summarize, our work contributes to the broader literature on Web privacy by utilizing computer vision to comprehensively capture aesthetic manipulation. Our findings show that prior works underestimate the prevalence of aesthetic manipulation. We attempt to explain the variance in salience between different buttons on the same banner as a function of banner design elements, and identify button placement as a new form of aesthetic manipulation. Furthermore, we collect data from EU and non-EU websites from EU and US IP addresses to determine the effects of user and website location on cookie banner design. This leads to our important finding that EU websites (which are most affected by GDPR) are more innovative in their cookie banner designs, and find legal ways to nudge users towards sharing their personal data (e.g., paywalls and aesthetic manipulation). This shows the importance of privacy research as a mechanism for uncovering new banner designs to continually update privacy regulation and better protect user privacy.}

\yc{I suggest to have a summary of contributions in a single paragraph or as bullets. I always try to label a paper as "first" of doing sth, if applicable. I suggest to have the findings listed, with actual numbers, better numbers in the literature and the numbers that we found. also summarize the trend and changes that we found. the affect of location}
\section{Related Work}
\yc{I will revisit this section later, but a general comment, it is too long. it should be 0.5-0.75 page}

There are two areas of research which are related to our paper: 1) studies of the prevalence and compliance of cookie banners on the Web and 2) studies which try to determine the effectiveness of different dark pattern designs. 

\subsection{Prevalence of Cookie Banners and Dark Patterns}

There are three studies that are similar to ours. The first related work manually visited 389 German websites one year apart and determined the ease of opting out and the prevalence of aesthetic manipulation~\cite{berens}. 
The second paper builds a system to automatically detect cookie banners and identify ten forms of non-compliant/dark pattern designs with high accuracy on 10,000 websites~\cite{kirkman}. The third paper visits 911 websites from both US and EU IP addresses in the 18 months following GDPR enforcement and records whether websites show a banner and how that banner hides the reject button~\cite{warberg}. Importantly, this is the first study to employ a SOD model to directly quantify button salience and identify cases of aesthetic manipulation, moving beyond prior approaches that focuses solely on button highlighting. We further differentiate ourselves by identifying new banner design categories (e.g., reject button placed in the corner). Only the third paper considers how the website and user location affects banner implementation, but aesthetic manipulation is not considered, and there are only five banner design categories. This means our paper is the most detailed study to investigate how website and user location affect banner designs, and the first to investigate how they affect aesthetic manipulation. 

 Other prior work has studied the prevalence of cookie banners and their designs since the implementation of GDPR. Published articles in this area between 2019 and 2024 show a generally positive trend for both the prevalence and compliance of cookie banners (see Table~\ref{tab:related} in Section~\ref{subsec:results_comp}).
 One such study uses longitudinal data to directly study how the percentage of websites with a cookie banner changes after the implementation of GDPR~\cite{degeling2019}. Other studies use cross-sectional data to investigate cookie banner compliance~\cite{utz,kampanos2021,Nouwens2020,Matte2020}. We report specific findings of these articles in Section~\ref{subsec:results_comp} when comparing our findings to results from the prior literature. We differentiate ourselves from these related works by investigating cookie banner designs for websites outside of the EU and studying how compliance varies based on the user's IP address. Furthermore, none of these related works examines aesthetic manipulation or a broad range of banner designs. Some additional prior work studies the prevalence of cookie paywalls~\cite{morel22, stenwreth}, but this is just one of the 17 cookie banner categories that we consider. 

There are related works (in addition to the aforementioned~\cite{kirkman}) which use machine learning and rule-based algorithms to automatically determine whether cookie banners and/or dark patterns exist on websites. Automated tools are beneficial in that they can scale to much larger data, but most current methods suffer from poor performance. Key performance metrics (i.e., accuracy, precision, and recall) often fall below 65\%, which we do not find acceptable. An exception is a tool called \textit{BannerClick}, which can detect cookie banners with almost perfect accuracy~\cite{rasaii}, but it is not capable of identifying the type of cookie banner design. 

\subsection{Dark Pattern Effectiveness}

There is prior work that determines what design patterns in cookie banners count as dark patterns. The authors of these studies predefine the design patterns they want to test and then manufacture a set of cookie banners to reflect the variables they want tested. The manufactured banners are then shown to the experiment's participants, and the users' consent elections are recorded. The authors then use a combination of acceptance rates from the experiment, and survey responses from post-experiment, to determine which cookie banner designs count as dark patterns. 
Some of the considered factors are types of aesthetic manipulation, including highlighting the accept button~\cite{bermejo}, button size~\cite{bauer}, replacing buttons with links, and banner location~\cite{habib}.

However, there are issues with using these types of experiments to identify forms of aesthetic manipulation. A survey of dark patterns in cookie banners cautions that many of these experiments suffer from small sample sizes and their users being habituated to select ``accept'' due to the proliferation of cookie banners that do not give users a real choice on the Web~\cite{bielovareport}. If users agree to personal data sharing out of habit, it makes the acceptance rate a less meaningful measure of the affect of aesthetic manipulation. Additionally, we believe that allowing personal data sharing in a controlled experiment may be incompatible with the users' real choices because the participants think that their data will be protected by the researchers. Finally, we think that it is hard to capture the complexity of aesthetic manipulation in a way that is testable in an experiment. For example, one study considers a highlighted accept button with a blue background~\cite{utz}. They do not consider different highlight colors, button text colors, or reject button formats. Thus, we conclude that the banners in the experiments do not reflect banners that are used by websites. We address the aforementioned issues for the aesthetic manipulation dark pattern by calculating and comparing the salience of each button directly to determine if aesthetic manipulation is present, instead of indirectly measuring it through the consent elections of study participants. Furthermore, we run our study on cookie banners that are actually used by websites instead of cookie banners that are manufactured for an experiment. 

\section{Method}
\label{sec:method}

Our analysis is performed on images of cookie banners from a large set of websites. Although there are a few datasets of cookie banner images~\cite{kirkman,berens}, these existing datasets do not fit our needs. Firstly, these images in existing datasets were captured only from a single point of view (i.e., an EU IP address). We intend to explore the differences between cookie banners on websites when visited from areas where privacy regulation exists (e.g., the EU) versus areas where privacy regulation does not exist (e.g., New York). Secondly, those images were not recently captured (both cited papers collect data in 2022) and thus do not reflect the current state of cookie banners on the Web. In particular, the old datasets would not reflect guidance from the European Data Protection Board released in January 2023~\cite{noyb}. Given these limitations, we collect the data ourselves. 

After data collection, our analysis consists of two parts. The first part is a manual analysis to determine banners' compliance with existing regulations and recommendations. The second part of our analysis utilizes a SOD tool to determine the relevance of each button on the cookie banner. We present the data collection process, the manual labeling method, and the SOD method in detail in the following three subsections. 

%

\subsection{Data Collection}
\label{sec:collection}

\begin{figure}[t!]
\centering
\includegraphics[width=0.9\linewidth]{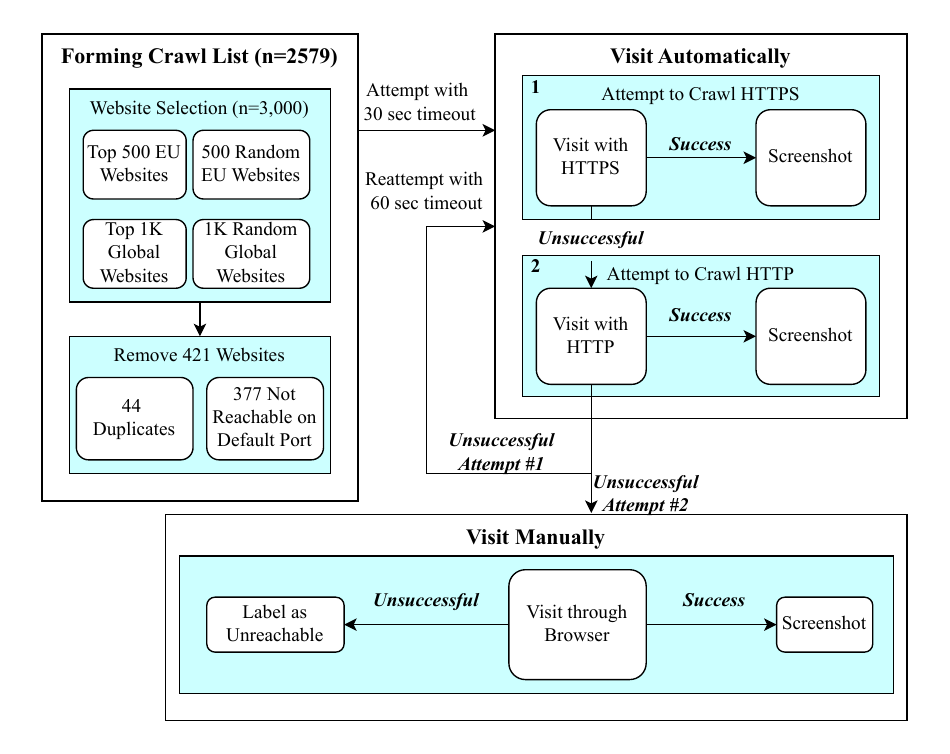}
\caption{Data Collection Process} 
\label{fig:collection}
\end{figure}

Our data collection process is shown in Figure~\ref{fig:collection}. We first create a list of websites that we are interested in visiting. We chose to study cookie banner implementations on popular and unpopular websites, as well as websites based inside and outside of the EU. Thus, our target domain list can be split into four categories: (1) popular global websites, (2) popular EU websites, (3) random global websites, and (4) random EU websites. 
\yc{I feel pieces is informal, and changed it to categories.}

\begin{table*}[t!]
\resizebox{\textwidth}{!}{
\begin{tabular}{|>{\centering\arraybackslash}m{0.12\textwidth}|>{\centering\arraybackslash}m{0.43\textwidth}|>{\centering\arraybackslash}m{0.45\textwidth}|}
\hline
\textbf{Label} & 
\textbf{Defining Characteristic} & 
\textbf{Laws/Recommendations} 
\\
\hline
\makecell{Notice}  & \makecell{Banner has no option to reject (or manage) cookies} & \makecell{\textbf{Not Compliant:} no choice to opt-out~\cite{gdpr}} \\
\hline
\makecell{Paywall}  & \makecell{Required to subscribe (pay) or allow cookies} & \makecell{\textbf{Likely Compliant:} no unified stance exists, but most\\ DPAs, including the CNIL in France, allow when \\the price is fair~\cite{morel22} } \\
\hline
\makecell{Full*}  & \makecell{Banner has accept and reject buttons, and some button\\ to set detailed privacy preferences (on another page) } & \makecell{\textbf{Compliant}\\~\cite{cnilrecs,noyb}} \\
\hline
\makecell{Full choices*}  & \makecell{Has accept and reject buttons, and allows \\for setting detailed privacy preferences on the banner} & \makecell{\textbf{Compliant}\\~\cite{cnilrecs,noyb}} \\
\hline
\makecell{Choices*}  & \makecell{Can set detailed privacy preferences on the banner, \\but missing at least one of the accept or reject buttons} & \makecell{\textbf{Likely Compliant:} have the ability to opt-in or \\opt-out on first page for each individual purpose, but we\\ are unaware of specific guidance on this banner type} \\
\hline
\makecell{Manage*}  & \makecell{There is no option to directly reject cookies on the\\ banner, but you can do so through settings} & \makecell{\textbf{Not Compliant:} too difficult to opt-out \\~\cite{cnilrecs,noyb}} \\
\hline
\makecell{Full-Manage*}  & \makecell{Banner only has accept and reject buttons} & \makecell{\textbf{Likely Not Compliant:} do not allow separate consent\\ choice to be made for each data processing\\ purpose individually~\cite{santos2020banners}
} \\
\hline
\makecell{Corner Reject}  & \makecell{The only way to reject cookies is to click a button\\ located outside of the banner or in the upper corner} & \makecell{\textbf{Likely Compliant:} same options as ``Full'' banner, but\\
we
are unaware of specific guidance on this banner type} \\
\hline
\makecell{Settings Only}  & \makecell{No accept or reject button. Only a button that takes\\ you to detailed settings and an ``X'' to close it.} & \makecell{\textbf{Likely Not Compliant:} unclear if user is consenting\\ when selecting ``X''~\cite{kirkman} } \\
\hline
\makecell{Preselected}  & \makecell{Presents detailed privacy preferences on the banner,\\ but the default is opt-out instead of opt-in} & \makecell{\textbf{Not Compliant:} result of the European Court of \\Justice's ruling in Planet 49 case~\cite{planet49}} \\
\hline
\makecell{Ambiguous}  & \makecell{Does not contain standard cookie banner options\\ (e.g., ``Do Not Sell''option)} & \makecell{\textbf{Likely Not Compliant:} no option to opt-out of data\\collection (can only opt-out of data being sold)} \\
\hline
\makecell{Two Banners}  & \makecell{Multiple cookie banners appear on the landing page} & \makecell{\textbf{Likely Not Compliant:} possibly violating the \\``informed and unambiguous" consent requirement\\~\cite{kirkman}, and two banners \\ means neither is "clearly distinguishable''~\cite{gdpr}} \\
\hline
\end{tabular}}
\caption{Categories of Cookie Banners and Compliance with Laws and Recommendations}
\caption*{\centering Note: * means that there is an additional category with the same defining characteristics in addition to an ambiguous ``X'' button on the banner. It is unclear if clicking the ``X'' results in implicit consent, which makes all such cases \textbf{Likely Not Compliant}~\cite{kirkman}.}
\label{tab:classes}
\end{table*}

\yc{I prefer to not use the turquoise color, or any color, in the boxes, less contrast may impose challenges to visually impaired readers}
We use the Tranco rankings~\cite{tranco} to determine which websites are popular\footnote{We use the Tranco list from December 12th, 2024: \url{https://tranco-list.eu/list/PNPLJ/1000000}}. The Tranco rankings are designed as an objective ranking of popular web domains to be used in academic research. For categories (1) and (3), respectively, we select the top 1,000 ranked websites and then randomly sample 1,000 additional websites from the Tranco list. For categories (2) and (4), we first filter the Tranco list based on the countries of the websites, as indicated by each website's country code top-level domains (ccTLDs)~\cite{degeling2019, Matte2020}.  Specifically, we only keep websites whose ccTLD matches one of the codes for the EU (.eu), the United Kingdom (.uk), or one of the 27 member states of the EU (e.g., .be for Belgium). We choose to include the United Kingdom because they still implement the GDPR (now called the UK GDPR), even after leaving the EU. 
We select the top 500 ranked websites and then randomly sample 500 additional websites from the filtered list to capture websites in categories (2) and (4), respectively.

All website visits were made between December 16 and December 23, 2024. Although the websites of all four categories total 3,000 websites, 44 websites appear on both the list of the top 1,000 global and top 500 EU websites and are visited only once. We also removed 377 websites that we found were not reachable on the regular HTTP/HTTPS port numbers, when tested with Socket\footnote{\url{https://docs.python.org/3/library/socket.html}} in Python. This leaves us with 2,579 websites. 

We build a crawler, using the Selenium\footnote{\url{https://www.selenium.dev/documentation/}} package in Python, that can automatically visit the websites in our target list and screenshot the landing page. 
As shown in Figure~\ref{fig:collection}, the crawler is initially run with a 30-second timeout and attempts to visit the website using Secure Hypertext Transfer Protocol (HTTPS). If successful, a screenshot of the landing page is saved. If there is an error, the crawler will retry accessing the website using Hypertext Transfer Protocol (HTTP). If the error persists after trying to visit with HTTPS and HTTP, we will retry the process with a longer timeout of 60 seconds. If still unsuccessful, we resort to manually visiting the website and will screenshot the landing page (if the website is reachable). 
\yc{I am curious how many manual visits are there}

We attempt to visit every target domain twice using virtual private networks (VPNs), once from an IP address in the EU (Paris, France), and the second time from an IP address in New York (NY). We chose New York because the proposed comprehensive data privacy regulation, the New York Privacy Act, has yet to be passed and signed into law, despite being introduced in 2021~\cite{NYPA}. Of the 2,579 websites in the list, we obtain screenshots of landing pages for 2,490 websites when crawling from Paris, and 2,492 websites when crawling from New York. The difference is due to the fact that some websites block access based on the location of the IP address.
\yc{why only Paris and France in EU? "an area without comprehensive privacy regulation" refers to NY only, right? if so, it should belong to the next sentence}



\subsection{Manual Labeling}
\label{subsec:labeling}
Our first research goal is to study the compliance of website cookie banner implementations with existing regulations and recommendations. Each captured image is manually labeled by a rater (one of the authors) as unreachable (different from the standard port issue), having no cookie banner, or in one of the 17 categories of Table~\ref{tab:classes} based on the ``Defining Characteristic'' listed. 
Some images are labeled unreachable since the websites block or discourage 
automated crawlers (resulting in a CAPTCHA or ``Access Denied'' page). Also, some websites have notifications and advertisements blocking the cookie banner.
%
In these cases, we visit the website manually to validate the label. If the manual visit is successful, we take a new screenshot and relabel the website. 

For quality control, each image is labeled twice to ensure high accuracy in the manual labeling process. \newtext{In the case of differing labels, the website was manually visited, and the rater interacted with the banner to determine the true label.} If a website receives different labels for visits from EU and NY IP addresses, we review the images for the third time. We load both images at the same time and ensure that the differing labels are due to actual differences in the landing page/cookie banner. \newtext{We further recruited two Ph.D. students (one with prior research involving cookie banners and one without) as raters to label 200 randomly sampled websites. We found consensus among the three raters was very high, with a Krippendorff's alpha of 0.94. Disagreements were re-evaluated by actually visiting the website and interacting with the banner. The verification shows all the labels made by the initial rater were correct.} 

We use these final categorizations to analyze the compliance of websites' cookie banners with the existing ``Laws/Recommendations'' shown in Table~\ref{tab:classes}. These categorizations are also used to differentiate the images that are applicable for SOD analysis.

\begin{figure}[t!]
\centering
\includegraphics[width=0.75\linewidth]{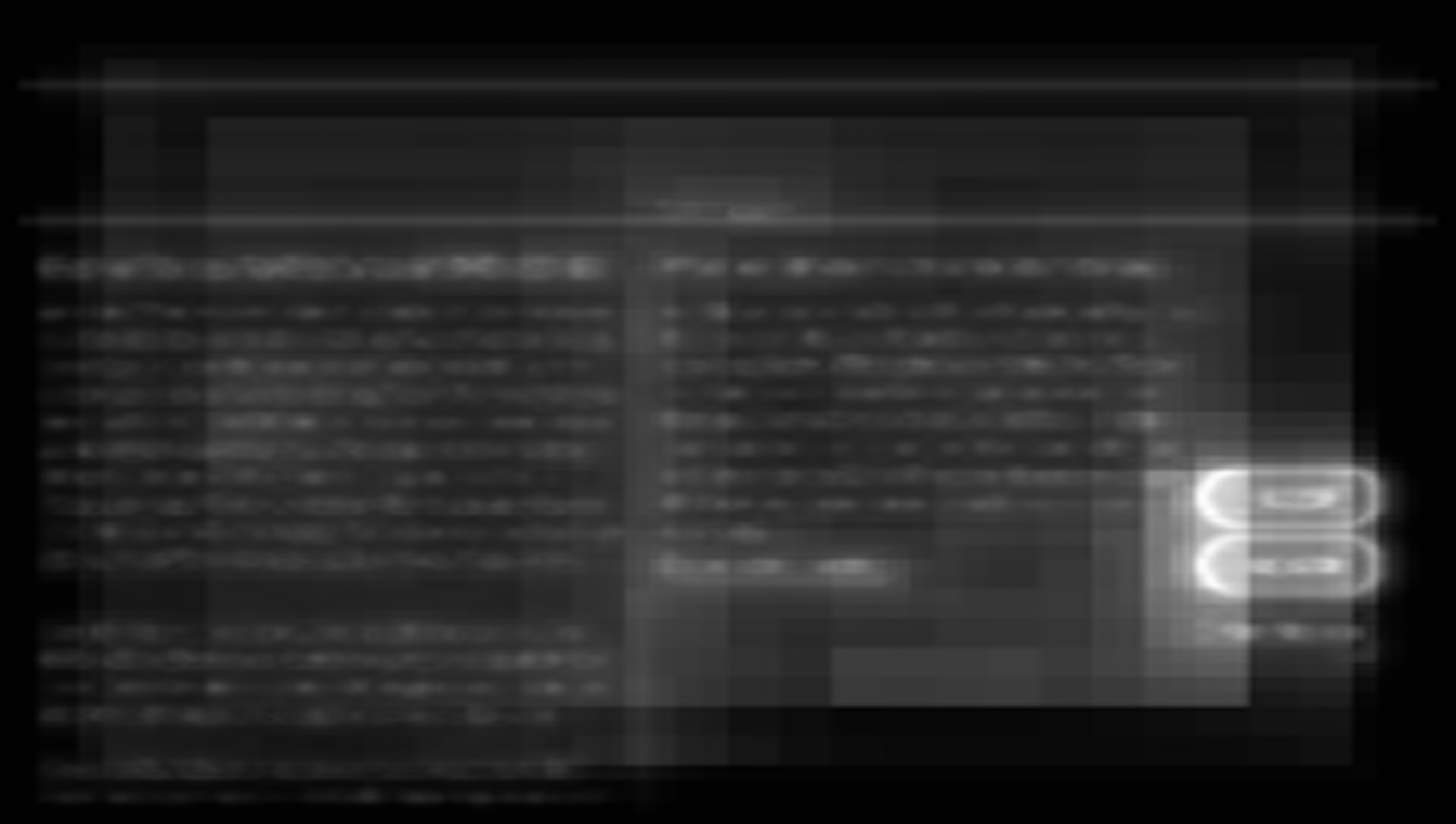}
\caption{\centering Visualization of Salient Areas for Landing Page of \textit{washingtonpost.com}} 
\label{fig:grayscale}
\end{figure}

\subsection{Salient Object Detection (SOD)}
\label{subsec:sod}
For the cookie banners that have multiple buttons, we want to compare the salience of each button. Thus, we separate out all websites that were manually labeled with a cookie banner in the categories ``Full'', ``Full choices'', ``Choices'', ``Manage'', ``Full-Manage'', or ``Corner Reject'' (including those with an ambiguous ``X'' button to close the banner). We then take the screenshots of the landing pages of these websites and pass the image to an SOD tool called \textit{DeepRare}~\cite{dr19,dr21} to measure the salience of each button.  We then compare the salience of the ``accept'' button (i.e. the button that allows the website to collect personal data with user consent) with the salience of other buttons. 
\yc{a bit confused with above sentence. Does it correspond to a graph or table? exact definition? how is it related to the next sentence}

We choose DeepRare because it does not require any additional training to be applied to our data. This eliminates the need for costly eye-tracking studies to label the areas of banners that are most salient to users. DeepRare has performed well when applied to various training datasets, and generalizes better than other models~\cite{dr21}. 
DeepRare is generalizable without training because of the authors' intuition that trained DNN-based SOD models learn to detect top-down objects (e.g., humans) instead of bottom-up features (e.g., colors). Thus, the authors conclude that other DNN-based SOD models actually learn to detect objects that are commonly considered as salient instead of learning to detect saliency itself~\cite{dr19,dr21}. DeepRare instead relies on the detection of ``rarity'' within the features extracted by pretrained convolutional neural networks (CNNs). \newtext{Details are provided in Appendix~\ref{app:dr}.} The output of DeepRare is a matrix of pixel-level salience scores the same size as the input image. The salience scores for each pixel are between 0 and 255, where higher scores mean the corresponding pixel is more salient. This matrix of values can be visualized as a grayscale image. The brightest areas of the grayscale image indicate the salient areas of the input image (e.g., Figure~\ref{fig:grayscale} shows the salient pixels for the landing page of \textit{washingtonpost.com}).

\begin{figure}[t!]
\centering
\includegraphics[width=0.8\linewidth]{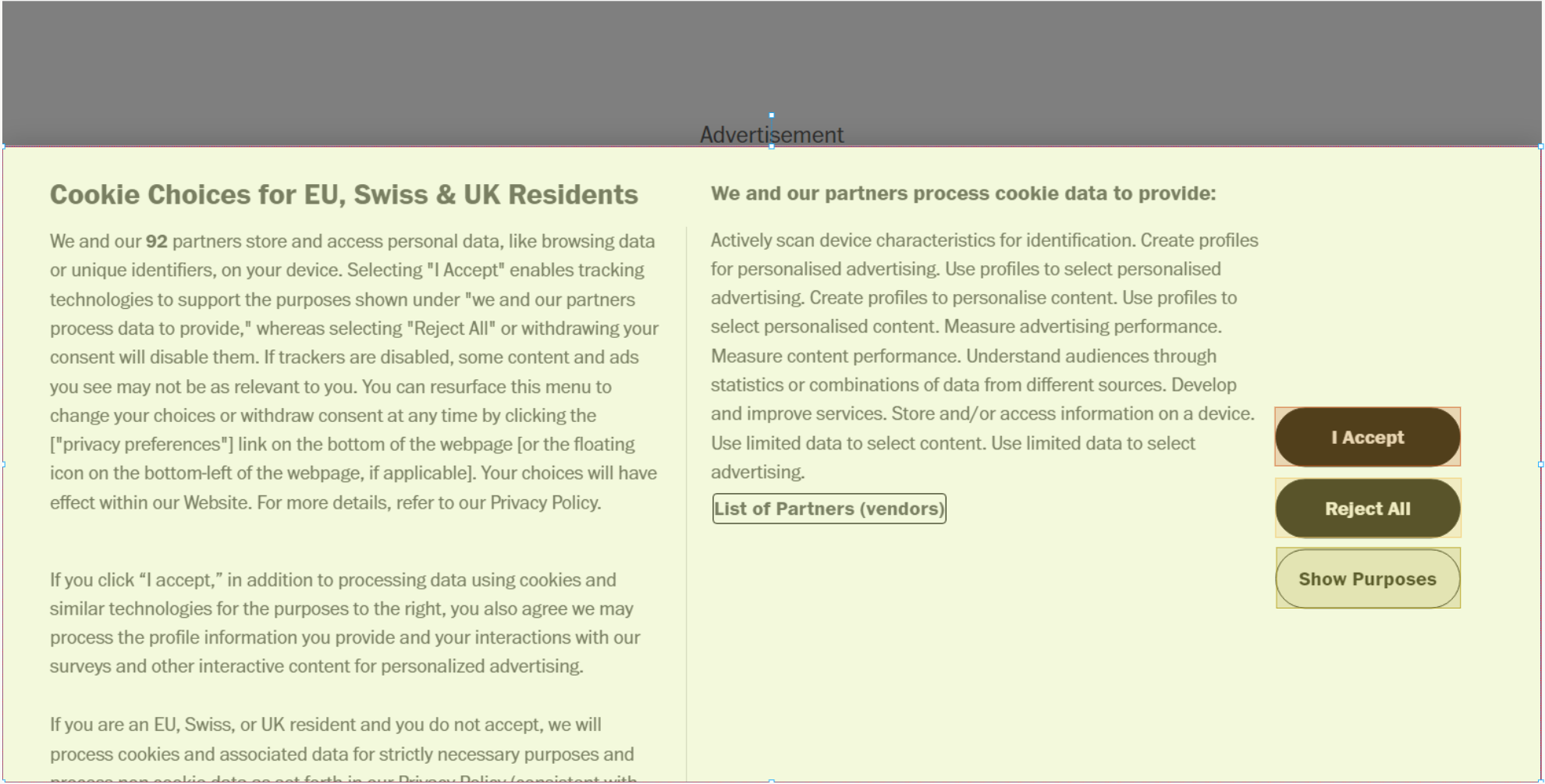}
\caption{\centering Labeling Cookie Banner Elements for \textit{washingtonpost.com}} 
\label{fig:labelstudio}
\end{figure}

 Since DeepRare outputs pixel-level salience scores, we need to map each cookie banner element (e.g., buttons) to a defined range of pixels so that we can measure the salience of the element. We use an open source data-labeling tool called \textit{Label Studio}\footnote{\url{https://labelstud.io/}} to manually determine the pixel ranges of cookie banners and each of their elements with bounding boxes. We show an example image with bounding boxes after labeling the cookie banner, accept button, reject button, and manage settings button in Figure~\ref{fig:labelstudio}.

\yc{are you using "element" and "button" interchangeably? if so, is ist better to have a single term? if not, describe the differences for the first time, when the second term appears}

 To calculate the salience of the banner element, we can either use the average or maximum salience value of all pixels within the bounding box of the banner element. In the \textit{washingtonpost.com} example shown in Figures~\ref{fig:grayscale} and~\ref{fig:labelstudio}, the average salience values are 161.6, 160.9, and 84.5 for the accept, reject, and manage settings buttons, respectively. The maximum salience values are 255, 252, and 169, respectively. From either measure, the accept and reject buttons are clearly more salient than the manage settings button (likely due to the darker color). On the other hand, the salience values of the accept and reject buttons are very similar. It is hard to know if these minor differences are due to subtle dark patterns (e.g., positioning the accept button on top) or noise.
 
 For the purposes of this paper, we choose to take a conservative approach \newtext{and label dark patterns only when we can confidently conclude the score difference is due to true differences in button salience. Towards this goal, we first perturb the original input image by injecting small amounts of noise (e.g., slightly changing the pixel hues or flipping the picture over its vertical/horizontal axes). We describe this process in more detail in Appendix B, but the goal is to reduce the impact of noisy DeepRare outputs on the results. After injecting noise, we calculate the salience scores for each pixel in the perturbed images and average these scores together to get a more robust measure of salience.}
 
\newtext{We borrow two methods from salient object ranking literature to transform pixel-level salience scores into button-level scores. The bounding box labeling denotes which pixels belong to which button, and we either take the average or maximum score of all the pixels belonging to a button as the button's salience score~\cite{kalashsalience,liusalience}. These methods capture different characteristics of salience (i.e., the single most salient pixel may be different from the most salient region of pixels), but we show in Appendix C that the trends are the same (i.e., the accept button is the most salient). For this reason we choose to use the sum of the normalized average and normalized maximum salience scores as our measure of button salience.} 
 
\newtext{To determine if a button's salience score indicates a true difference in salience, we define a threshold based in psychophysics. Prior research shows that visual salience follows Weber's Law, which states that the smallest detectable change in stimulus (i.e., the just noticeable difference (JND)) is proportional to the original stimulus value~\cite{huang2005}. This means a percentage value can be calculated to represent the minimum threshold at which a change in stimulus becomes perceptible. Although this value has not been experimentally calculated for visual salience due to the difficulty in measuring discrete changes in a subjective stimulus, Weber's law holds for many visual tasks (e.g., contrast, angular/objective size). The empirically calculated JNDs in these visual tasks range from 1.6\% in contrast to 6.8\% in angular size~\cite{pelli2013,mckee1992}. Thus, we choose a conservative threshold value of 7\%, and only consider aesthetic manipulation to occur if the ``accept'' button is 7\% more salient than the other buttons. We do present aesthetic manipulation frequency at different thresholds in Appendix D.} 

\section{Results}

We present our results in five parts: 1) an analysis of GDPR compliance, the prevalence of each type of cookie banner design from Table~\ref{tab:classes}, and how user/website location affects these results; 2) a comparison to prior work that categorizes cookie banners on the Web; 3) an analysis of button salience on cookie banners and the prevalence of aesthetic manipulation; 4) a comparison to prior work that reports the prevalence of aesthetic manipulation; and 5) identification of new forms of aesthetic manipulation.
\yc{"location of the website", not location of user?}

\subsection{Cookie Banner Type and GDPR Compliance}
~\label{subsec:results_comp}

We present our findings broken down by the location of the website and the IP address of the visitor, which is a novel contribution of our work. We choose to show frequencies as a percentage of websites or banners because the sample sizes in each of the four groups are different. We successfully reached 1,009 EU websites from an EU IP address, 1,010 EU websites from a US IP address, 1,308 non-EU websites from an EU IP address, and 1,335 non-EU websites from a US IP address. We first investigate how many of the banner designs are noncompliant with GDPR and related EU privacy rulings (e.g., Planet 49) and recommendations (e.g., data protection authorities)~\cite{planet49,cnilrecs}. 
Using the categorizations of each banner type as compliant/non-compliant or likely compliant/non-compliant (from the ``Laws/Recommendations'' column of Table~\ref{tab:classes}), we can determine the percentage of (likely) compliant banners. We report the percentage of banners in each combination of website and visitor location in Table~\ref{tab:compliance}.  

\begin{table}[t!]
\resizebox{\linewidth}{!}{
\begin{tabular}{|>{\centering\arraybackslash}m{0.235\linewidth}|>{\centering\arraybackslash}m{0.18\linewidth}|>{\centering\arraybackslash}m{0.195\linewidth}|>{\centering\arraybackslash}m{0.195\linewidth}|>
{\centering\arraybackslash}m{0.195\linewidth}|}
\hline
\textbf{Location} & 
\textbf{Compliant} & 
\textbf{Likely Compliant} & 
\textbf{Likely Not Compliant}&
\makecell{\textbf{Not}\\ \textbf{Compliant}}
\\
\hline
\makecell{\textbf{EU Site}\\ \textbf{EU Visitor}} & 44.9\% & 17.0\% & 12.2\% & 25.9\% \\
\hline
\makecell{\textbf{EU Site}\\ \textbf{US Visitor}} & 40.9\% & 16.7\% & 15.0\%& 27.4\% \\
\hline
\makecell{\textbf{non-EU Site}\\ \textbf{EU Visitor}} & 48.4\% & 2.2\% & 19.9\%& 29.4\%  \\
\hline
\makecell{\textbf{non-EU Site}\\ \textbf{US Visitor}} & 13.5\% & 1.2\% & 22.3\%& 63.0\%\\
\hline
\end{tabular}
}
\caption{Compliance by Website and Visitor Location}
\label{tab:compliance}
\end{table}

Table~\ref{tab:compliance} shows that websites within the EU or visited from an EU IP address are more likely to show a compliant cookie banner. In such a case, the percentage of fully compliant banners is greater than 40\% and the percentage of noncompliant banners is less than 30\% (the remaining $\sim$30\% of banners are either in the likely compliant or likely non-compliant categories). 
%
On the other hand, if both the visitor and the website are outside the EU, the percentage of compliant banners falls to 13.5\% and the percentage of noncompliant banners skyrockets to 63.0\%. This finding is to be expected as GDPR and data protection authority guidance is not applicable unless the data processor (i.e., the website) or the data subject (i.e., visitor) is in the EU. 

\yc{the earlier sentence uses whole number, but here and some other places have 2 decimals, make them consistent}

\begin{figure}[t!]
\centering
\includegraphics[width=0.9\linewidth]{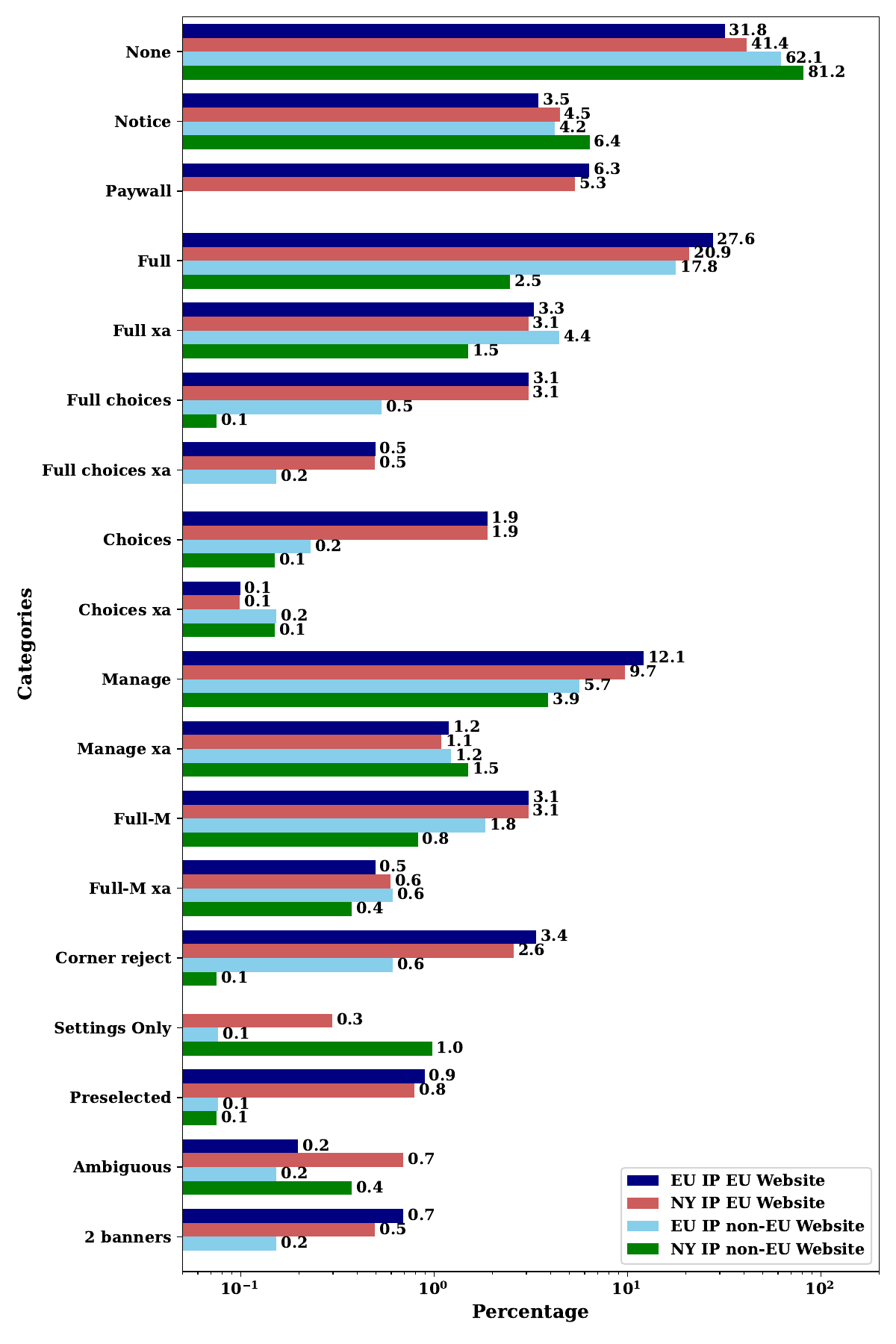}
\caption{Frequency of Each Cookie Banner Type (logarithmic scale)}
\label{fig:percentage_categories}
\end{figure}

Next we look at the likely compliant and likely non-compliant categories. Their categorization is determined by whether the design still follows key requirements for valid consent in GDPR. For example, a cookie paywall is likely compliant because it still gives users a choice between opting-in/out. On the other hand, a design without a detailed settings option is likely not compliant because it does not give a user an option to consent to each data-sharing purpose individually. As shown in Table~\ref{tab:compliance}, EU websites are more likely to implement cookie banners that are classified as likely compliant, while non-EU websites are more likely to implement banners classified as likely non-compliant. 
This shows how EU websites, \newtext{that} are familiar with GDPR \newtext{and heavily affected by it, are able to come up with innovative cookie banner designs}\deltext{and ensure their designs follow key requirements, while also using this familiarity} to nudge users towards sharing personal data without violating the law. This suggests that websites will continually exploit banner designs that are not explicitly outlawed to \textit{legally} push users toward sharing personal data. Indeed, we were unable to find unified and specific legal guidance from data protection authorities in the EU on many categories of cookie banner designs (ones that are classified as likely compliant/not compliant in this study). Unless more resources are put into web privacy research and regulation, innovative websites that are hungry for more user data will continue to create new cookie banner designs before they are addressed by regulation.

Figure~\ref{fig:percentage_categories} reports the percentage of websites with each type of cookie banner design, including a category titled ``None'' (i.e., no banner is shown on the website). As we can see, the absence of any banner on a website is more likely when either the website or visitor is located outside the EU. Thus, cookie banners are more frequent and more compliant (as shown in Table~\ref{tab:compliance}) when the user and/or the website are in the EU. \newtext{The absence of a cookie banner is not necessarily a violation of GDPR as it is unnecessary if the website only uses cookies that are essential for functionality. It is beyond the scope of this paper to determine website compliance in the absence of a cookie banner.}

There are two unexpected findings shown in Figure~\ref{fig:percentage_categories}. First, we see that cookie paywalls are popular on EU websites (6.3\% and 5.4\% of EU websites for EU and US visitors, respectively), but are nonexistent on non-EU websites. One possible explanation is that websites prefer to collect users' personal data instead of collecting a fixed price per user, and thus, only implement cookie paywalls if the number of users opting out is high. This implies that the value of user data to a website may be higher than anticipated.

Second, we see very different results for EU websites visited from EU versus US IP addresses. For instance, 27.6\% of websites show a ``Full'' banner to an EU visitor, but only 20.9\%  of them show a ``Full'' banner to a US visitor. This is surprising because GDPR still applies to EU websites even when the visitor is from outside the EU~\cite{gdpr}. Thus, an EU website should not change their cookie banner design based on a visitor's location. 


\begin{figure}[t!]
\centering
\includegraphics[width=0.8\linewidth]{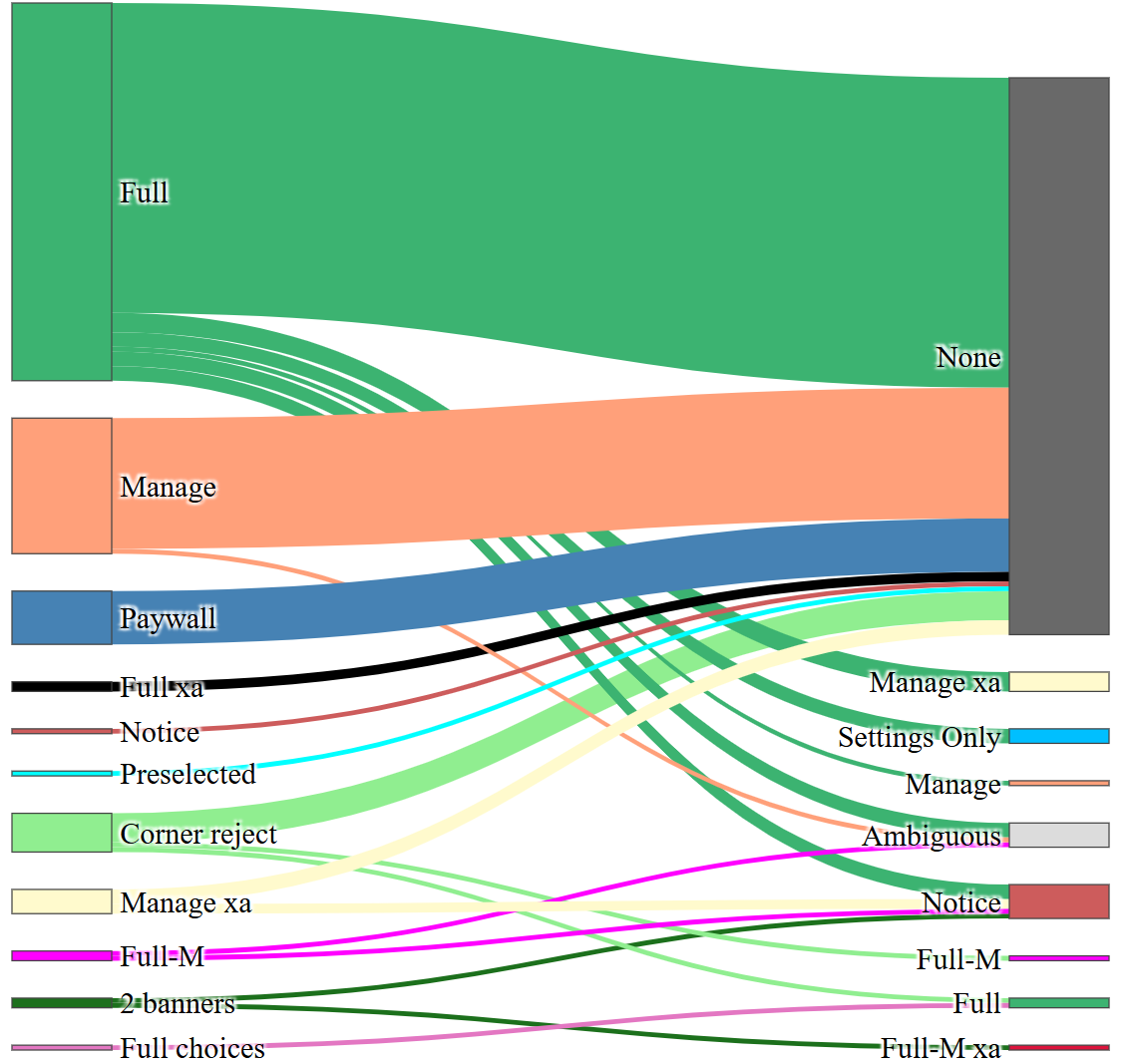}
\caption{\centering How EU Websites Change Banner Design when Visitor is from the US}
\label{fig:sankey}
\end{figure}

\begin{table*}[t!]
\resizebox{\textwidth}{!}{
\begin{tabular}{|>{\centering\arraybackslash}m{0.04\textwidth}|>{\centering\arraybackslash}m{0.09\textwidth}|>{\centering\arraybackslash}m{0.06\textwidth}|>
{\centering\arraybackslash}m{0.07\textwidth}|>{\centering\arraybackslash}m{0.03\textwidth}|>{\centering\arraybackslash}m{0.12\textwidth}|>{\centering\arraybackslash}m{0.12\textwidth}|>{\centering\arraybackslash}m{0.09\textwidth}|>{\centering\arraybackslash}m{0.09\textwidth}|>
{\centering\arraybackslash}m{0.06\textwidth}|>{\centering\arraybackslash}m{0.1\textwidth}|>
{\centering\arraybackslash}m{0.05\textwidth}|>{\centering\arraybackslash}m{0.08\textwidth}|}
\hline
\textbf{ID}&
\textbf{Year} & 
\textbf{Size} & 
\textbf{Website} &
\textbf{IP} & 
\textbf{\% with any banner} & 
\textbf{\% with opt-out choice} &
\textbf{\% Accept + Reject} & 
\textbf{\% Manage} & 
\textbf{\% Full-M} &
\textbf{\% Preselected} & 
\textbf{\% xa} &
\textbf{\% 2 Banners} 
\\
\hline
0& 2018& 1,000& EU& EU& - & 14.0\% & 3.2\% & 9.8\% & - & - & - & - \\
\hline
1& 2018-2018& 4,125& EU& EU& 50.0\%-68.6\% & - & - & - & - & - & - & - \\
\hline
1& 2018-2018& 2,462& non-EU& EU& 25.9\%-44.6\% & - & - & - & - & - & - & - \\
\hline
2 & 2018-2019& 467& EU& EU& $\sim$75\% - $\sim$80\% & $\sim$40\% - $\sim$75\% & - & - & - & - & - & - \\
\hline
2 & 2018-2019& 444& non-EU& EU& $\sim$35\% - $\sim$35\% & $\sim$50\% - $\sim$67\% & - & - & - & - & - & - \\
\hline
2 & 2018-2019& 467& EU& US& $\sim$70\% - $\sim$60\% & $\sim$43\% - $\sim$67\% & - & - & - & - & - & - \\
\hline
2 & 2018-2019& 444& non-EU & US& $\sim$10\% - $\sim$10\% & $\sim$10\% - $\sim$25\% & - & - & - & - & - & - \\
\hline
3 & 2019& 680& EU& EU& - & 67.5\% & 11.8\% & - & - & 56.2\% & - & - \\
\hline
4 & 2019& 560& EU & EU& - & 93.2\% & - & - & - & 46.5\% & - & - \\
\hline
5 & 2020& 17,737& EU & EU& 45\% & 70\% & 8\% & 62\% & 3\% & - & - & - \\
\hline
6 & 2021-2022& 389& EU & EU& 65.6\% - 76.9\% & 71.0\% - 85.8\% & 15.3\% - 47.8\% &55.7\% - 38.1\% & - & - & - & - \\
\hline
7 & 2022& 10K& Mixed & EU& 47\% & - & - & - & - & - & - & - \\
\hline
7 & 2022 & 10K & Mixed & US & 30\% & - & - & - & - & - & - & - \\
\hline
8 & 2022 & 10,127 & Mixed & EU & 23.9\% & 66.2\% & 32.4\% & 33.8\% & - & 1.9\% & 17.5\% & 7.9\% \\
\hline
\hline
Ours & 2024& 1,009 & EU & EU& 68.2\% & 94.9\% & 60.6\% &22.4\% & 5.2\% & 1.3\% & 8.2\% & 1.0\% \\
\hline
Ours & 2024 & 1,308 & non-EU & EU& 37.9\% & 88.9\% & 68.5\% & 19.4\% & 6.4\% & 0.2\% & 17.5\% & 0.4\% \\
\hline
Ours & 2024 & 1,010 & EU & US & 58.6\% & 92.4\% & 57.6\% & 23.1\% & 6.2\% & 1.3\% & 9.6\% & 0.9\% \\
\hline
Ours & 2024 & 1,335 & non-EU & US & 18.8\% & 66.1\% & 28.2\% & 35.5\% & 9.9\% & 0.4\% & 23.9\% & 0\% \\
\hline
\end{tabular}}
\caption{Comparison to Related Work}
\caption*{\centering Mapping ID to Citation: 0:~\cite{utz}, 1:~\cite{degeling2019}, 2:~\cite{warberg}, 3:~\cite{Nouwens2020}, 4:~\cite{Matte2020}, 5:~\cite{kampanos2021}, 6:~\cite{berens}, 7:~\cite{rasaii}, 8:~\cite{kirkman}}
\label{tab:related}
\end{table*}

Upon further investigation, we find that among the 1,003  EU websites that are reachable from US and EU IP addresses, 139 of them (13.9\%) have a cookie banner design for EU visitors and either no banner or a different design for US visitors. We illustrate these design changes with a Sankey diagram in Figure~\ref{fig:sankey}. The left column shows the banner design when a visitor is from the EU, and the right column shows the design of the same websites when the visitor is from the US. Among the 139 websites that have changed the designs, 115 (82.7\%) of them completely remove their banner and 14 (10.1\%) replace a fully compliant banner with a less compliant one (e.g., can not opt-out or can only opt-out through the settings page). Of the 115 websites that completely remove their banner, 64 (55.7\%) show a fully compliant banner to EU visitors. We are unaware of prior research that reports any similar finding. 

\yc{be consistent throughout the paper, whole number, single decimal, double decimal}


\subsubsection{Comparison to Prior Findings.}
We also compare our findings with related work (Table~\ref{tab:related}) to place our findings in the evolving landscape of cookie banner design and its relation to GDPR compliance. We compare with any work that has a sample size of more than 300 websites and does not rely on automated systems with poor accuracy. Before our work, the most recent data collection was in 2022. After that, new guidance on banners~\cite{noyb} and fines against non-compliant banners have occurred. Our work provides evidence as to whether such guidance and enforcement are effective in changing industry standards. This work is the second study (and the first was conducted in 2018-2019) that investigates the four possible combinations of website/visitor location inside/outside of the EU~\cite{warberg}.

\yc{"more than 300 website" could have a large upper bound. can you say about sth like 300, less than 400 websites, about 300-350 websites, instead? need to cite the first study. also, is banner configuration same as banner design, if so, use the same term}

Historically, the two most studied questions regarding cookie banners are: 1) the percentage of websites that display a cookie banner, and 2) the percentage of those banners that users can opt out of. It is difficult to identify a trend for the first question because the results are heavily influenced by the type of websites studied. Early studies focus on popular websites~\cite{degeling2019,warberg}, while later studies include more unpopular websites~\cite{kampanos2021,kirkman}. Unpopular websites are less likely to have a banner because they are less likely to collect personal data and need a banner, and they are also more likely to escape GDPR enforcement due to a smaller number of people being affected. However, one thing has been clear throughout all the studies: more websites display cookie banners to EU visitors than US visitors~\cite{warberg, rasaii}, and EU visitors will find more banners on EU websites than non-EU websites~\cite{degeling2019,warberg}. Our findings confirm this result. The second question has easily identifiable trends. For US visitors, the percentage of banners that could be opted out of increased by 15-25 percentage points between 2018 and 2019~\cite{warberg}. Our findings align with this as US visitors can opt-out of 66.1\% and 88.9\% of the banners on non-EU and EU websites, respectively, which are the highest reported percentages.

\deltext{Now, we compare our results for visiting websites with an EU IP address to the existing work.}

\deltext{Looking at all of the studies in Table~\ref{tab:related} shows that}\newtext{Similarly,} the percentage of banners that can be opted out of for EU visitors has increased over time, with our study reporting the highest percentages (88.9\% and 94.9\% of cookie banners on non-EU and EU websites, respectively). Fewer articles study the other types of banner designs shown in Table~\ref{tab:related}, and none of them visits websites with a US IP address, despite EU websites being required to follow GDPR regardless of the user's location~\cite{gdpr}. Our study fills in this gap, but for now, we compare our results for visiting websites with an EU IP address to the existing work. 

First, since 2018, the percentage of banners with an accept and reject button on the first page has increased. Between 2021 and 2022, this number increased drastically (15\% to 48\%)~\cite{berens}. This is likely due to CNIL fines against Meta and Google in December 2021 for making it easier to opt-in than to opt-out~\cite{cnilfines}. This trend has continued in our findings, as the percentage of banners with accept and reject buttons on the first page is higher in our dataset than in any previous dataset (60.6\% of EU websites and 68.5\% of non-EU websites). As the percentage of banners with a reject button on the first page increases,  the percentage of banners only allowing the user to opt-out through settings is expected to decrease. We see this trend in previous results and it continues in our findings (the ``\% Manage'' column in Table~\ref{tab:related}). A study in 2021~\cite{kampanos2021} reported that 3\%  of banners have accept and reject buttons, but without setting options. We report a higher percentage of this banner design: 5.2\%, which may represent the increased attention to adding a rejection button following the CNIL fines. 

Second, banner designs with preselected options were frequent (46.5\%-56.2\%) in 2019~\cite{Matte2020,Nouwens2020}, but had significantly fallen to 1.9\% of banners by 2022~\cite{kirkman}. This is likely due to the Planet 49 ruling where the CJEU ruled such designs to be non-compliant. We find that designs with preselected options remain infrequent in 2024 (under 2\% of the banners). 

Finally, we compare our results for the percentage of banners with an ambiguous close option (the ``\% xa'' column) and percentage of websites with two cookie banners to a paper with 2022 data~\cite{kirkman}. The percentage of banners with an ambiguous close option is consistent, but we find significantly fewer websites with two banners ($\sim$1\% vs 7.9\%). It is not clear why this decrease in prevalence occurred.

In general, several important trends are established in Table~\ref{tab:related}. First, when a ruling states that a banner design is noncompliant with GDPR, most websites are quick to comply (e.g., no preselected options, or need a reject button on the first page of the banner). However, when no such ruling exists, questionable banner designs can become popular (e.g., preselected options before the Planet 49 ruling, or the ambiguous close option currently). Second, while US web users are not specifically protected by GDPR, they still see some benefits (e.g., increased banner prevalence and increased ability to opt-out of sharing personal data). 

Although not put in the table, there are a couple of studies that specifically study the prevalence of cookie paywalls. 
We identified cookie paywalls on 6.3\% of EU websites when visited from an EU IP address, a significant increase from 0.7\% of 2,800 websites in the first study of cookie paywalls~\cite{morel22}. A recent paper visits websites that are known to have cookie paywalls from IP addresses inside and outside the EU and finds they are more frequently shown to EU visitors~\cite{stenwreth}. This is consistent with our findings (6.3\% versus 5.4\% of EU websites when visited from the EU and US, respectively.).

\subsection{Salience Analysis and Aesthetic Manipulation}

Aesthetic manipulation is a dark pattern in cookie banner design where the accept button is designed to be more prominent than the other buttons (specifically, the reject button)~\cite{gray}. Our goal is to show that in addition to the non-compliant banner designs from Section~\ref{subsec:results_comp}, many banners also use aesthetic manipulation to nudge users toward opting-in. Two related works investigated the prevalence of aesthetic manipulation~\cite{berens,kirkman}, but both relied on a single aspect: the number of websites that highlight the accept button. As we show, there are other ways to make the accept button more salient, such as button position and type. To get a more complete picture of aesthetic manipulation, we use a computer vision model for SOD, called DeepRare, to directly measure the salience of each button. 

\deltext{We measure the salience of a button with two methods borrowed from the literature in the area of SOD and salient object ranking~\cite{kalashsalience,liusalience}. The salience of a button can either be calculated as the average or maximum salience score of all pixels that are part of the button. We use histograms to show the frequencies of normalized salience scores for the accept, reject, and manage buttons on all \newtext{compliant} banners using the average and maximum methods in Figure~\ref{fig:avgsal}. The figure shows that at lower scores, the most frequent button is manage, followed by reject, and then accept. At higher scores, this ordering reverses, and accept buttons are most frequent, with manage buttons being the least frequent. Both measures of salience show this trend, but we believe that they capture slightly different characteristics of salience \newtext{(i.e., the single most salient pixel may be different from the most salient region of pixels)}. For this reason, we chose to combine the two measures into a single salience metric. For each button, we consider the salience score to be \newtext{the sum of the normalized average salience score and normalized maximum salience score, and use this summed score for the remaining analysis. }\deltext{a weighted sum of the two scores where the weight is chosen so that each contribute equally (weighted sum because maximum salience score is always higher than average salience score). We use this weighted score for the remaining analyses.}}

We focus on the \newtext{894} cookie banners having accept, reject, and manage buttons that were previosuly classified as compliant (i.e., Full, Full Choices, or Corner Reject). There could be aesthetic manipulation on other banner designs, but we consider it less important as these designs already have more overt manipulation tactics to nudge a user towards opting-in (e.g., hiding the reject button in settings). \newtext{As shown in Table~\ref{tab:salience}, 38.0\% of the compliant banners have aesthetic manipulation (i.e., the accept button is significantly more salient than reject and manage buttons). In comparison, only 7.6\% and 6.2\% of compliant banners make the reject button or manage button, respectively, the most salient button. We attribute the majority of these instances to incidental aesthetic manipulation (e.g., a choice menu is large and salient, but probably is not intended to nudge users towards privacy-friendly options). This incidental aesthetic manipulation may occur in favor of the accept button too, thus, showing why guidelines should be updated to clearly state what is allowed in cookie banner design. The remaining 48.2\% of banners do not have one button that is substantially more salient than the others.}

\newtext{Another interesting takeaway from Table~\ref{tab:salience} is that aesthetic manipulation in compliant banners occurs 48.3\% more frequently on EU websites (42.4\%, $n=611$) than non-EU websites (28.6\%, $n=283$). A likely explanation is that EU websites have more incentives to use aesthetic manipulation, as they are forced to show banners to all users and cannot use more explicit manipulation tactics (e.g., not offer a reject button).}
 
 \newtext{Surprisingly, aesthetic manipulation in compliant banners is 28.5\% more likely to occur when the visitor is from NY (44.6\%, $n=303$) than the EU (34.7\%, $n=591$). This seemingly contradicts the first finding because we would expect more explicit manipulation tactics for non-EU users. A possible explanation is that websites only showing banners to EU users do not employ aesthetic manipulation because cookie banners are only shown to a subset of their users (especially for non-EU websites). }

\deltext{As shown in Table~\ref{tab:salience}, 
the percentage of compliant banners where the accept buttons is at least 5\% more prominent than the other two buttons is 40.7\%, and the percentage for the manage button is 14.3\%, and for the reject button is 6.3\%. The remaining 38.7\% of the banners do not have a single button that is at least 5\% more prominent than the others (we discuss the selection of a 5\% threshold in the appendix). Thus, the most common design pattern for a compliant banner is one that employs aesthetic manipulation to nudge users toward selecting ``accept''.} 

\deltext{There are other takeaways from Table~\ref{tab:salience}. Aesthetic manipulation is 33.7\% more likely to take place on EU websites (44.4\%) than non-EU websites (33.2\%). It is a somewhat surprising finding, as it shows that stringent privacy regulation does not always result in more privacy friendly implementations. A likely explanation for this is that EU websites have more incentives to use aesthetic manipulation as they are forced to show banners to all users and cannot use more explicit manipulation tactics (e.g., not offer a reject button). }

\deltext{Aesthetic manipulation is also 25.9\% more likely to occur when the visitor's IP address is in New York (47.1\%) than in the EU (37.4\%). This seemingly contradicts the first finding. A possible explanation is that websites that only show a banner to EU users are less likely to use aesthetic manipulation because cookie banners only inhibit them from collecting data from a small proportion of their visitors. Websites that show a banner to all users may be more determined to nudge users toward opting-in. Since EU users see more banners overall, it dilutes the percentage of banners with aesthetic manipulation. }

\begin{table}[t!]
\resizebox{\linewidth}{!}{
\begin{tabular}{|>{\centering\arraybackslash}m{0.16\linewidth}|>{\centering\arraybackslash}m{0.26\linewidth}|>{\centering\arraybackslash}m{0.32\linewidth}|>{\centering\arraybackslash}m{0.26\linewidth}|}
\hline
\multicolumn{4}{|c|}{\textbf{Accept}} \\ 
\hline
\textbf{Location} & 
\textbf{EU Website} & 
\textbf{non-EU Website} & 
\textbf{Total} \\
\hline
\textbf{EU IP} & 39.9\% & 27.4\% & 34.7\%  \\
\hline
 \textbf{NY IP} & 45.5\% & 37.1\% & 44.6\%\\
\hline
\textbf{Total} & 42.4\% & 28.6\% & 38.0\%\\
\hline
\hline
\multicolumn{4}{|c|}{\textbf{Reject}} \\ 
\hline
\textbf{Location} & 
\textbf{EU Website} & 
\textbf{non-EU Website} & 
\textbf{Total} \\
\hline
\textbf{EU IP} & 6.1\% & 8.1\% & 6.9\%  \\
\hline
 \textbf{NY IP} & 8.6\% & 11.4\% & 8.9\%\\
\hline
\textbf{Total} & 7.2\% & 8.5\% & 7.6\%\\
\hline
\hline
\multicolumn{4}{|c|}{\textbf{Manage}} \\ 
\hline
\textbf{Location} & 
\textbf{EU Website} & 
\textbf{non-EU Website} & 
\textbf{Total} \\
\hline
\textbf{EU IP} & 5.2\% & 7.7\% & 6.3\%  \\
\hline
 \textbf{NY IP} & 5.2\% & 11.4\% & 5.9\%\\
\hline
\textbf{Total} & 5.2\% & 8.1\% & 6.2\%\\
\hline
\end{tabular}
}
\caption{\centering Percentage of Compliant Banners where the Button is 7\% more Salient than other Buttons}
\label{tab:salience}
\end{table}

\subsubsection{Comparison to Prior Findings.} We can compare our results with the findings of previous studies on the prevalence of aesthetic manipulation. However, both studies report the prevalence of aesthetic manipulation differently than we do in Table~\ref{tab:salience}, choosing to focus on all banners with accept and reject buttons, regardless of GDPR compliance. \newtext{Both methods only report aesthetic manipulation when button highlighting is present.~\citet{berens} choose to manually label their dataset to identify button highlighting, but~\citet{kirkman} automate the identification with grayscale button contrast. We use this method from~\citeauthor{kirkman} to compare our findings with previous work. Our proposed method identifies aesthetic manipulation in 38.0\% of all compliant banners in our sample, while the method from~\citeauthor{kirkman} only identifies aesthetic manipulation in 27.0\% of the same banners. A McNemar's test for paired nominal data confirms this difference is statistically significant $(\chi^2(1)=57.51, p<.001)$. This supports our belief that prior work underestimates the prevalence of aesthetic manipulation due to their singular focus on button highlighting.}\deltext{Both existing studies collect data in 2022 and visit the websites from an EU IP address. One finds aesthetic manipulation on 51.6\% of EU websites~\cite{berens} and the other finds aesthetic manipulation on 26\% of 1,600 websites (a mix of EU and non-EU websites)~\citeauthor{kirkman}. When visiting from an EU IP address, we find aesthetic manipulation on 57.64\% and 43.6\% of the EU and US websites with an accept and reject button, respectively. and reject button use aesthetic manipulation when visited from an EU IP address. Thus, our results show a higher prevalence of aesthetic manipulation.}

\deltext{An additional way to compare our results is to use the method of identifying aesthetic manipulation in~\cite{kirkman} that identifies button highlighting through grayscale button contrast. Using this method identifies aesthetic manipulation in 25.6\% of the compliant banners in our sample, instead of the 40.7\% that we identified (from Table~\ref{tab:salience}). This further shows that prior methods, which only consider button highlighting, underreport the prevalence of aesthetic manipulation.}

\subsubsection{Result Significance}

\newtext{We are interested in determining if our findings of aesthetic manipulation, and the effects of user/website locations, are statistically significant. We first test whether the button label (i.e., accept, reject, or manage) is significantly correlated with the button salience in the 894 compliant banners. A repeated measures ANOVA with the Greenhouse-Geisler correction (due to violation of the sphericity assumption) reveals that the button label has a significant effect on button salience $(F(1.965,1786)=474.51,p<.001,\eta_{g}^2=0.167)$. Due to violating the assumption of normality, we use the Wilcoxon Signed Rank Test with the Bonferroni correction as the post hoc pairwise test to show that the accept button is significantly more salient than the manage button $(W=29313,p<.001)$ and the reject button $(W=89371,p<.001)$. These results support our findings in Table~\ref{tab:salience}: websites with compliant banners often use aesthetic manipulation to make the accept button more salient.}


\newtext{We are also interested in how website location, visitor IP address location, and their interactions with button label affect salience. Thus, we regress button salience on button label, IP address location, website location, and their interactions. We include website-level fixed effects because we only want to explain the button salience variance within the same cookie banner (i.e., we do not want factors such as website layout or presence of background images to affect our results). The results of this regression can be found in Table~\ref{tab:regression_results_main}, and there are two main takeaways. First, the manage button is consistently less salient than the accept button (as indicated by a statistically significantly negative coefficient for the more statistically significant negative coefficient of the manage button label), while the reject button is only significantly less salient on EU websites. The lack of statistical significance on any term including EU IP address indicates that aesthetic manipulation is largely driven by website location, and not visitor location. Second, cookie banners on EU websites tend to have more salient buttons, probably because of the emphasis on explicit consent in GDPR. If the user makes no consent election, the website can not collect/process the user's data, so websites make the entire banner more salient to encourage a consent election.}

\begin{table}[t]
\centering
\begin{tabular}{l c}
\hline
\textbf{Variable} & \textbf{Coefficient (SE)} \\
\hline
Intercept &  1.321 (0.04)\textsuperscript{***} \\
Reject Button & -0.081 (0.061) \\
Manage Button & -0.252 (0.061)\textsuperscript{***} \\
EU website & 0.152 (0.042)\textsuperscript{***}\\
EU IP Address & -0.045 (0.146)\\
Reject x EU website & -0.185 (0.065)\textsuperscript{***}\\
Reject x EU IP & -0.014 (0.066)\\
Manage x EU website & -0.117 (0.065) \\
Manage x EU IP & -0.087 (0.066)\\
EU website x EU IP & 0.081 (0.147)  \\
Reject x EU IP x EU website & 0.034    
(0.072)  \\
Manage x EU IP x EU website & 0.026 (0.072)\\
\hline
\multicolumn{2}{r}{\textsuperscript{***}$p<0.01$}
\end{tabular}
\caption{\centering Salience Regressed onto Button Label, Website Location, and Visitor Location}
\label{tab:regression_results_main}
\end{table}

\subsection{Identifying New Forms of Aesthetic Manipulation}

\newtext{A main contribution of our method is that it is not limited to only identifying aesthetic manipulation in the presence of button highlighting.We find aesthetic manipulation to be more prevalent than previously reported, and want to identify the other design elements that contribute to aesthetic manipulation.}\deltext{To identify other aesthetic manipulation dark patterns, we first measure the importance of a button as a function of the characteristics of the button and banner.} For the \newtext{1,621} banners with at least two buttons, we collect the following \newtext{design} elements \newtext{for investigation}. 

\begin{itemize}
    \item \textit{Button Size}: area in pixels
    \item \textit{Brightness}: average grayscale value of button pixels
    \item \textit{Contrast}: absolute difference between button and banner brightness
    \item \textit{Button Distance}: distance (in pixels) between center of element and center of page
    \item \textit{Button to Banner Distance (BB Distance)}: distance (in pixels) between center of button and center of banner
    \item \textit{Corner}: button is in the banner's upper corner 
    \item \textit{Link}: the button is formatted as a link 
    \item \textit{Hidden}: the button is hidden in the banner text 
    \item \textit{Choice Menu}: the button is actually a choice checklist
\end{itemize}

\begin{table}[t]
\centering
\begin{tabular}{l c}
\hline
\textbf{Variable} & \textbf{Coefficient (SE)} \\
\hline
Intercept &  0.684 (0.062)\textsuperscript{***} \\
Button Size & 0.007 (0.003)\textsuperscript{**} \\
Brightness & -0.010(0.003)\textsuperscript{***}\\
Contrast & 0.105 (0.003)\textsuperscript{***}\\
Button Distance & -0.084 (0.004)\textsuperscript{***}\\
BB Distance & 0.022 (0.004)\textsuperscript{***} \\
Corner & -0.157 (0.009)\textsuperscript{***}\\
Link & -0.074 (0.007)\textsuperscript{***} \\
Hidden & -0.023 (0.016) \\
Choice Menu & -0.051 (0.016)\textsuperscript{***} \\
\hline
\multicolumn{2}{r}{\textsuperscript{***}$p<0.01$, 
\textsuperscript{**}$p<0.05$ 
}
\end{tabular}
\caption{Button Salience Regressed onto Design Choices}
\label{tab:regression_results}
\end{table}

\newtext{We do not use the perturbed images for this analysis, because the noise injections are similar to some of the variables of interest (e.g., changing image brightness).} After collecting these elements for each button, we run a least squares regression with website fixed effects to identify how these characteristics affect salience. \newtext{We again include website fixed effects because we do not want webpage variation to affect our results (e.g., one website may have a large, salient, company logo on their page that makes all banner elements less salient).} \newtext{For interpretability and comparability,} all non-binary variables were standardized \newtext{to have a mean of 0 and standard deviation of 1,} and the salience score was normalized so that values fall between 0 and 1. The regression results are shown in Table~\ref{tab:regression_results}. 


The \newtext{statistically significant negative coefficient for the }brightness variable indicates that darker buttons are more prominent, which is logical as most of the content on the landing page of these websites is white, empty space. The \newtext{large, statistically significant positive coefficient for the }contrast variable indicates that the contrast between the button and the banner is one of the strongest indicators of high salience. This is related to previous studies that measure aesthetic manipulation only as a highlighted button. 

 We are unaware of any past research that considers button locations, but our regression shows that this is important. Intuitively, we find that buttons located closer to the center of the webpage are more salient. However, buttons near the center of the cookie banner are less salient (although the relationship is less strong). This can probably be explained by the fact that the banner text is located in the center, so buttons closer to the center are more likely to be crowded by text. However, moving a button all the way to the corner also has a negative relationship with salience.

\newtext{Button size has a positive, statistically significant relationship with button salience. This finding is intuitive, and the coefficient's magnitude may be weakened by the presence of the binary indicators for corner buttons, hidden buttons, and link buttons, because all of these can overlap with decreased button size.} Finally, displaying a button as a choice menu or link has a significant, negative relationship with salience. Displaying a button as a link likely decreases salience because the links are often smaller than buttons and look similar to other text. \deltext{The fact that links are the only frequent use of button size manipulation may explain why the button size variable is not significant.} \newtext{Although past research has considered whether using links or smaller reject buttons could nudge users towards opting-in~\cite{bauer}, it has never been considered in studies that detect aesthetic manipulation on the Web.}

These variables, with website-fixed effects, account for 83.5\% of the variance in salience scores. A regression that only considers brightness and contrast with website fixed-effects explains 77\%, \newtext{and fixed effects alone explain 51.8\%} of the variance in salience scores. This further shows that our method of detecting aesthetic manipulation is more comprehensive than simply considering button highlighting.

\section{Conclusion}
This study provides a comprehensive analysis of the compliance of cookie banners with GDPR and the prevalence of aesthetic manipulation, revealing that \newtext{38.0\%} of the compliant banners employ aesthetic manipulation to nudge users toward consent. Our analysis of 2,579 websites, visited from EU and US IP addresses, highlights an alarming trend: EU-based websites have higher compliance rates, but are also more likely to utilize aesthetic manipulation. This suggests that privacy regulation and its enforcement, while increasing compliance, may unintentionally incentivize the development of more subtle forms of manipulation.

These findings underscore the need for regulation that deals with dark patterns. By publicly releasing our dataset and methodology, we aim to contribute to ongoing research and policy discussions focused on mitigating these manipulative practices in an effort to increase user privacy on the Web. Future work should prioritize the development of improved detection mechanisms for other dark patterns (e.g., coercive or privacy shaming text). Our dataset could be used to generate ground truth saliency maps of cookie banners through eye-tracking studies. Although costly, such a dataset would allow the application of supervised SOD models for the task of detecting aesthetic manipulation. \newtext{Additionally, the dataset could be used to train a classifier than can identify the 17 cookie banner design categories from our study.} 
\section*{Acknowledgments}
This work was supported by the NSF under Grants No. CNS 2237328
and DGE 2043104. Thanks to Pritam Sen and Ahmad Alemari for their help with the manual labeling. 
\bibliography{aaai25}

\begin{thebibliography}{31}
\providecommand{\natexlab}[1]{#1}

\bibitem[{Bauer, Bergstrøm, and Foss-Madsen(2021)}]{bauer}
Bauer, J.~M.; Bergstrøm, R.; and Foss-Madsen, R. 2021.
\newblock Are you sure, you want a cookie? – The effects of choice architecture on users' decisions about sharing private online data.
\newblock \emph{Computers in Human Behavior}, 120.

\bibitem[{Berens et~al.(2024)Berens, Bohlender, Dietmann, Krisam, Kulyk, and Volkamer}]{berens}
Berens, B.~M.; Bohlender, M.; Dietmann, H.; Krisam, C.; Kulyk, O.; and Volkamer, M. 2024.
\newblock Cookie disclaimers: Dark patterns and lack of transparency.
\newblock \emph{Computers \& Security}, 136.

\bibitem[{Bermejo~Fernandez et~al.(2023)Bermejo~Fernandez, Chatzopoulos, Papadopoulos, and Hui}]{bermejo}
Bermejo~Fernandez, C.; Chatzopoulos, D.; Papadopoulos, D.; and Hui, P. 2023.
\newblock This Website Uses Nudging: MTurk Workers' Behaviour on Cookie Consent Notices.
\newblock In \emph{Proceedings of the ACM on Human-Computer Interaction}. ACM.

\bibitem[{Bielova(2022)}]{bielovareport}
Bielova, N. 2022.
\newblock {Survey of academic studies measuring the effect of dark patterns on acceptance consent rate of users in consent banners}.
\newblock Technical report, CNIL.

\bibitem[{CJEU(2019)}]{planet49}
CJEU. 2019.
\newblock Case C‑673/17: Planet49 GmbH v Bundesverband der Verbraucherzentralen und Verbraucherverbände – Verbraucherzentrale Bundesverband e.V.

\bibitem[{CNIL(2020)}]{cnilrecs}
CNIL. 2020.
\newblock Cookies and other trackers: the CNIL publishes amended guidelines and its recommendation.
\newblock \url{https://www.cnil.fr/fr/cookies-et-autres-traceurs/regles/cookies/lignes-directrices-modificatives-et-recommandation}.
\newblock Accessed: 2025-01-04.

\bibitem[{Degeling et~al.(2019)Degeling, Utz, Lentzsch, Hosseini, Schaub, and Holz}]{degeling2019}
Degeling, M.; Utz, C.; Lentzsch, C.; Hosseini, H.; Schaub, F.; and Holz, T. 2019.
\newblock We Value Your Privacy ... Now Take Some Cookies: Measuring the GDPR's Impact on Web Privacy.
\newblock In \emph{Proceedings of the 2019 Network and Distributed System Security Symposium}. Internet Society.

\bibitem[{EU(2016)}]{gdpr}
EU. 2016.
\newblock {Regulation (EU) 2016/679 of the European Parliament and of the Council of 27 April 2016 on the protection of natural persons with regard to the processing of personal data and on the free movement of such data (General Data Protection Regulation)}.
\newblock Official Journal of the European Union, L 119, 1--88.

\bibitem[{Gray et~al.(2018)Gray, Kou, Battles, Hoggatt, and Toombs}]{gray}
Gray, C.~M.; Kou, Y.; Battles, B.; Hoggatt, J.; and Toombs, A.~L. 2018.
\newblock The Dark (Patterns) Side of {UX} Design.
\newblock In \emph{Proceedings of the 2018 {CHI} Conference on Human Factors in Computing Systems}, 1--14. Association for Computing Machinery.
\newblock ISBN 978-1-4503-5620-6.

\bibitem[{Habib et~al.(2022)Habib, Li, Young, and Cranor}]{habib}
Habib, H.; Li, M.; Young, E.; and Cranor, L. 2022.
\newblock “Okay, whatever”: An Evaluation of Cookie Consent Interfaces.
\newblock In \emph{Proceedings of the 2022 CHI Conference on Human Factors in Computing Systems}. ACM.

\bibitem[{Huang and Pashler(2005)}]{huang2005}
Huang, L.; and Pashler, H. 2005.
\newblock Quantifying object salience by equating distractor effects.
\newblock \emph{Vision Research}, 45(14): 1909--1920.

\bibitem[{Kalash, Islam, and Bruce(2021)}]{kalashsalience}
Kalash, M.; Islam, M.~A.; and Bruce, N. D.~B. 2021.
\newblock Relative Saliency and Ranking: Models, Metrics, Data and Benchmarks.
\newblock \emph{IEEE Transactions on Pattern Analysis and Machine Intelligence}, 43(1): 204--219.

\bibitem[{Kampanos and Shahandashti(2021)}]{kampanos2021}
Kampanos, G.; and Shahandashti, S.~F. 2021.
\newblock Accept All: The Landscape of Cookie Banners in Greece and the UK.
\newblock In \emph{Proceedings of the ICT Systems Security and Privacy Protection}.

\bibitem[{Kirkman, Vaniea, and Woods(2023)}]{kirkman}
Kirkman, D.; Vaniea, K.; and Woods, D. 2023.
\newblock DarkDialogs: Automated detection of 10 dark patterns on cookie dialogs.
\newblock In \emph{Proceedings of the 2023 IEEE 8th European Symposium on Security and Privacy (EuroS\&P)}.

\bibitem[{Kong et~al.(2022)Kong, Mancas, Gosselin, and Po}]{dr21}
Kong, P.; Mancas, M.; Gosselin, B.; and Po, K. 2022.
\newblock DeepRare: Generic Unsupervised Visual Attention Models.
\newblock \emph{Electronics}, 11 (11): 1696.

\bibitem[{Le~Pochat et~al.(2019)Le~Pochat, Van~Goethem, Tajalizadehkhoob, and Joosen}]{tranco}
Le~Pochat, V.; Van~Goethem, T.; Tajalizadehkhoob, M., Samaneh~Korczy\'{n}ski; and Joosen, W. 2019.
\newblock Tranco: A Research-Oriented Top Sites Ranking Hardened Against Manipulation.
\newblock In \emph{Proceedings of the 26th Annual Network and Distributed System Security Symposium}, NDSS 2019.

\bibitem[{Liu et~al.(2022)Liu, Li, Zhao, Hao, and Shao}]{liusalience}
Liu, N.; Li, L.; Zhao, W.; Hao, J.; and Shao, L. 2022.
\newblock Instance-Level Relative Saliency Ranking With Graph Reasoning.
\newblock \emph{IEEE Transactions on Pattern Analysis and Machine Intelligence}, 44(11): 8321--8337.

\bibitem[{Mancas, Kong, and Gosselin(2020)}]{dr19}
Mancas, M.; Kong, P.; and Gosselin, B. 2020.
\newblock Visual Attention: Deep Rare Features.
\newblock In \emph{Proceedings of the 2020 Joint 9th International Conference on Informatics, Electronics \& Vision (ICIEV) and 2020 4th International Conference on Imaging, Vision \& Pattern Recognition (icIVPR)}.

\bibitem[{Matte, Bielova, and Santos(2020)}]{Matte2020}
Matte, C.; Bielova, N.; and Santos, C. 2020.
\newblock Do Cookie Banners Respect my Choice? : Measuring Legal Compliance of Banners from IAB Europe’s Transparency and Consent Framework.
\newblock In \emph{Proceedings of the 2020 IEEE Symposium on Security and Privacy (SP)}, 791--809. IEEE.

\bibitem[{McKee and Welch(1992)}]{mckee1992}
McKee, S.~P.; and Welch, L. 1992.
\newblock The precision of size constancy.
\newblock \emph{Vision Research}, 32(8): 1447--1460.

\bibitem[{Morel et~al.(2022)Morel, Santos, Lintao, and Human}]{morel22}
Morel, V.; Santos, C.; Lintao, Y.; and Human, S. 2022.
\newblock Your Consent Is Worth 75 Euros A Year - Measurement and Lawfulness of Cookie Paywalls.
\newblock In \emph{Proceedings of the 21st Workshop on Privacy in the Electronic Society}, 213--218. ACM.

\bibitem[{{New York State Senate}(2024)}]{NYPA}
{New York State Senate}. 2024.
\newblock Senate Bill S365B.
\newblock \url{https://www.nysenate.gov/legislation/bills/2023/S365/amendment/original/}.
\newblock Accessed: 2025-01-02.

\bibitem[{Nouwens et~al.(2020)Nouwens, Liccardi, Veale, Karger, and Kagal}]{Nouwens2020}
Nouwens, M.; Liccardi, I.; Veale, M.; Karger, D.; and Kagal, L. 2020.
\newblock Dark Patterns after the GDPR: Scraping Consent Pop-ups and Demonstrating their Influence.
\newblock In \emph{Proceedings of the 2020 CHI Conference on Human Factors in Computing Systems}, 1--13. ACM.

\bibitem[{{NOYB}(2024)}]{noyb}
{NOYB}. 2024.
\newblock Consent Banner Report: Overview of EU and national guidelines on dark patterns.
\newblock \url{https://noyb.eu/sites/default/files/2024-07/noyb_Cookie_Report_2024.pdf}.
\newblock Accessed: 2025-05-12.

\bibitem[{Pelli and Bex(2013)}]{pelli2013}
Pelli, D.~G.; and Bex, P. 2013.
\newblock Measuring contrast sensitivity.
\newblock \emph{Vision research}, 90: 10--14.

\bibitem[{Rasaii et~al.(2023)Rasaii, Singh, Gosain, and Gasser}]{rasaii}
Rasaii, A.; Singh, S.; Gosain, D.; and Gasser, O. 2023.
\newblock Exploring the Cookieverse: A Multi-Perspective Analysis of Web Cookies.
\newblock In \emph{Proceedings of the Passive and Active Measurement Conference (PAM ’23)}.

\bibitem[{Santos, Bielova, and Matte(2020)}]{santos2020banners}
Santos, C.; Bielova, N.; and Matte, C. 2020.
\newblock Are cookie banners indeed compliant with the law? : Deciphering EU legal requirements on consent and technical means to verify compliance of cookie banners.
\newblock \emph{Technology and Regulation}, 91--135.

\bibitem[{Stenwreth, Tang, and Morel(2024)}]{stenwreth}
Stenwreth, A.; Tang, S.; and Morel, V. 2024.
\newblock To Be or Not to Be (in the EU): Measurement of Discrepancies Presented in Cookie Paywalls.
\newblock \emph{arXiv preprint arXiv:2410.06920}.

\bibitem[{{The National Law Review}(2022)}]{cnilfines}
{The National Law Review}. 2022.
\newblock CNIL Fines Big Tech Companies 210 Million Euros for Cookie Violations.
\newblock \url{https://natlawreview.com/article/cnil-fines-big-tech-companies-210-million-euros-cookie-violations}.
\newblock Accessed: 2025-01-08.

\bibitem[{Utz et~al.(2019)Utz, Degeling, Fahl, Schaub, and Holz}]{utz}
Utz, C.; Degeling, M.; Fahl, S.; Schaub, F.; and Holz, T. 2019.
\newblock (Un)informed Consent: Studying GDPR Consent Notices in the Field.
\newblock In \emph{Proceedings of the 2019 ACM SIGSAC Conference on Computer and Communications Security}, 973--990.

\bibitem[{Warberg et~al.(2023)Warberg, Lefrere, Cheyre, and Acquisti}]{warberg}
Warberg, L.; Lefrere, V.; Cheyre, C.; and Acquisti, A. 2023.
\newblock Trends in Privacy Dialog Design after the GDPR: The Impact of Industry and Government Actions.
\newblock In \emph{Proceedings of the 22nd Workshop on Privacy in the Electronic Society}. ACM.

\end{thebibliography}

\subsection{Paper Checklist to be included in your paper}

\begin{enumerate}

\item For most authors...
\begin{enumerate}
    \item  Would answering this research question advance science without violating social contracts, such as violating privacy norms, perpetuating unfair profiling, exacerbating the socio-economic divide, or implying disrespect to societies or cultures?
    \answerYes{Yes}
  \item Do your main claims in the abstract and introduction accurately reflect the paper's contributions and scope?
    \answerYes{Yes}
   \item Do you clarify how the proposed methodological approach is appropriate for the claims made? 
    \answerYes{Yes, this is explained in Section~\ref{subsec:sod}, as well as in the Appendix.}
   \item Do you clarify what are possible artifacts in the data used, given population-specific distributions?
    \answerYes{Yes, we discuss that some websites block users depending on IP address, which means some websites are only reachable in one of our datasets (EU and US). }
  \item Did you describe the limitations of your work?
    \answerYes{Yes, see conclusion and Appendix D.}
  \item Did you discuss any potential negative societal impacts of your work?
    \answerNo{No, we do not anticipate how our work could have a negative societal impact. We identify patterns that are already hurting web users' privacy, so we hope that uncovering these patterns has a positive societal impact.}
      \item Did you discuss any potential misuse of your work?
    \answerNo{No, but we clarify that future work is needed in the conclusion.}
    \item Did you describe steps taken to prevent or mitigate potential negative outcomes of the research, such as data and model documentation, data anonymization, responsible release, access control, and the reproducibility of findings?
    \answerYes{Yes, we are releasing our dataset so the results can be reproduced and are releasing our code for data collection so the method can be tested on new data.}
  \item Have you read the ethics review guidelines and ensured that your paper conforms to them?
    \answerYes{Yes}
\end{enumerate}

\item Additionally, if your study involves hypotheses testing...
\begin{enumerate}
  \item Did you clearly state the assumptions underlying all theoretical results?
    \answerNA{NA}
  \item Have you provided justifications for all theoretical results?
    \answerNA{NA}
  \item Did you discuss competing hypotheses or theories that might challenge or complement your theoretical results?
    \answerNA{NA}
  \item Have you considered alternative mechanisms or explanations that might account for the same outcomes observed in your study?
    \answerNA{NA}
  \item Did you address potential biases or limitations in your theoretical framework?
    \answerNA{NA}
  \item Have you related your theoretical results to the existing literature in social science?
    \answerNA{NA}
  \item Did you discuss the implications of your theoretical results for policy, practice, or further research in the social science domain?
    \answerNA{NA}
\end{enumerate}

\item Additionally, if you are including theoretical proofs...
\begin{enumerate}
  \item Did you state the full set of assumptions of all theoretical results?
    \answerNA{NA}
	\item Did you include complete proofs of all theoretical results?
    \answerNA{NA}
\end{enumerate}

\item Additionally, if you ran machine learning experiments...
\begin{enumerate}
  \item Did you include the code, data, and instructions needed to reproduce the main experimental results (either in the supplemental material or as a URL)?
    \answerYes{Yes, details about reproducibility are included in the code/data repository.}
  \item Did you specify all the training details (e.g., data splits, hyperparameters, how they were chosen)?
    \answerNA{NA}
     \item Did you report error bars (e.g., with respect to the random seed after running experiments multiple times)?
    \answerNA{NA}
	\item Did you include the total amount of compute and the type of resources used (e.g., type of GPUs, internal cluster, or cloud provider)?
    \answerNo{No, because there are no special compute resources required as we do not have to train the model.}
     \item Do you justify how the proposed evaluation is sufficient and appropriate to the claims made? 
    \answerNA{NA}
     \item Do you discuss what is ``the cost`` of misclassification and fault (in)tolerance?
    \answerYes{Yes, we consider the cost of misclassification when trying to limit the affect of noisy outputs from DeepRare as described in Section~\ref{subsec:sod}.}
  
\end{enumerate}

\item Additionally, if you are using existing assets (e.g., code, data, models) or curating/releasing new assets, \textbf{without compromising anonymity}...
\begin{enumerate}
  \item If your work uses existing assets, did you cite the creators?
    \answerYes{Yes}
  \item Did you mention the license of the assets?
    \answerNo{No, because the authors stated the model was free to be adapted/used for research as long as they are cited.}
  \item Did you include any new assets in the supplemental material or as a URL?
    \answerYes{Our dataset and data collection code will be released in the published version of the paper (as a URL).}
  \item Did you discuss whether and how consent was obtained from people whose data you're using/curating?
    \answerYes{No, we are not dealing with any personal data.}
  \item Did you discuss whether the data you are using/curating contains personally identifiable information or offensive content?
    \answerNo{No, because we are not collecting data that relates to individuals.}
\item If you are curating or releasing new datasets, did you discuss how you intend to make your datasets FAIR?
\answerYes{Yes, we intend to make our dataset FAIR by uploading the data and detailed metadata to Zenodo.}
\item If you are curating or releasing new datasets, did you create a Datasheet for the Dataset)? 
\answerYes{Yes}
\end{enumerate}

\item Additionally, if you used crowdsourcing or conducted research with human subjects, \textbf{without compromising anonymity}...
\begin{enumerate}
  \item Did you include the full text of instructions given to participants and screenshots?
    \answerNA{NA}
  \item Did you describe any potential participant risks, with mentions of Institutional Review Board (IRB) approvals?
   \answerNA{NA}
  \item Did you include the estimated hourly wage paid to participants and the total amount spent on participant compensation?
    \answerNA{NA}
   \item Did you discuss how data is stored, shared, and deidentified?
   \answerNA{NA}
\end{enumerate}

\end{enumerate}

\appendix
\section{Details of DeepRare}
\label{app:dr}
\begin{figure}[t!]
\centering
\includegraphics[width=0.7\linewidth]{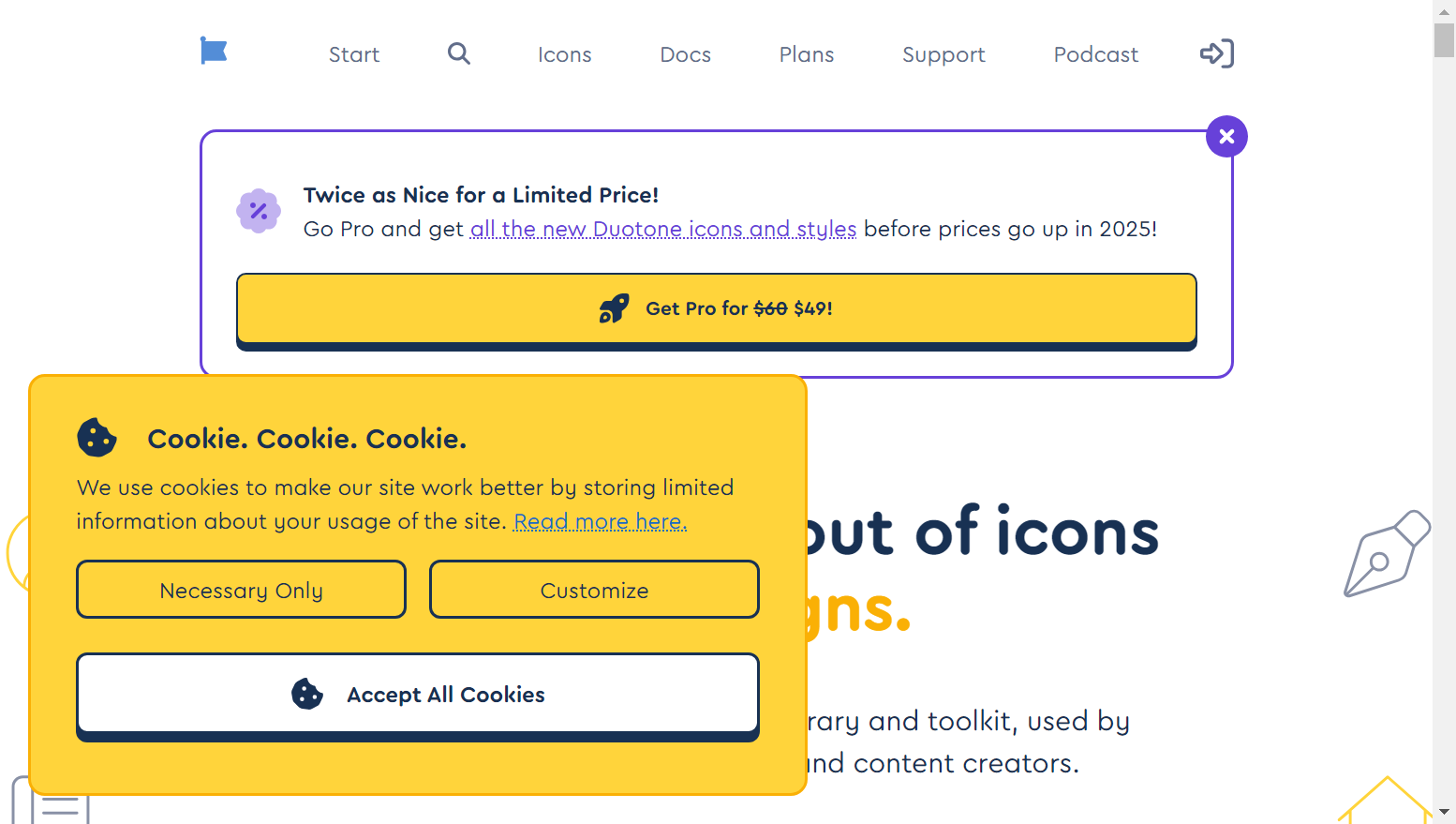}
\caption{\centering Original Input Image for \textit{fontawesome.com}}
\label{fig:fontawesome_orig}
\end{figure}
The version of DeepRare we use relies on a pretrained VGG16 model (pretrained on the popular ImageNet dataset). The pretrained VGG16 is capable of extracting important low-level, medium-level, and high-level features to understand images. The VGG16 architecture has 13 convolutional layers, which fall into five convolutional blocks which can be characterized by the type of features extracted: 1) the first two layers extract low-level features, 2) layers 4 and 5 extract higher low-level features, 3) layers 7,8, and 9 extract medium-level features, 4) layers 11,12, and 13 extract higher medium-level features, and 5) the last three convolutional layers extract high-level features~\cite{dr19,dr21}. The third, sixth, and tenth layers are pooling layers. Each of the convolutional layers produces a specific number of feature maps after convolving the input with the convolutional filters. Each convolutional layer in the block produces 64, 128, 256, 512, and 512 feature maps, respectively~\cite{dr21}. With all of the feature maps extracted, DeepRare takes four steps to calculate rarity. 


First, it calculates the rarity of the pixels at the level of the feature map. This is done for each feature map by dividing the values of the pixels and calculating $p(i)$, the probability of a pixel belonging to the bin $i$, for each bin. The rarity of all pixels in the bin is then defined as $R(i)=-\log p(i)$. The second step is to fuse rarity scores for each convolutional layer. For each pixel, the fused rarity score is a weighted sum of the rarity scores of that pixel in all feature maps produced by the convolutional layer. The weight of each pixel score in the sum is determined by the squared difference between the max and mean rarity scores for the feature map to which the pixel belongs. After the second step, there are 13 different feature maps that capture the salience of each pixel in the image determined by a single convolutional layer. The third step in DeepRare is to use the same fusion process to combine the weighted feature maps of all the layers in each convolutional block. This third step results in five feature maps, each showing the salience scores based on the level (e.g., low or high) of the extracted features. The fourth and final step of DeepRare is to sum and normalize (between 0 and 255) the salience scores of each pixel across the five different feature maps produced in the third step. The result of this final step is a matrix of values between 0 and 255 where the value $(i,j)$ is the pixel-level salience score for the corresponding pixel $(i,j)$ in the original image~\cite{dr19,dr21}.

\begin{figure}[t!]
\centering
\includegraphics[width=0.7\linewidth]{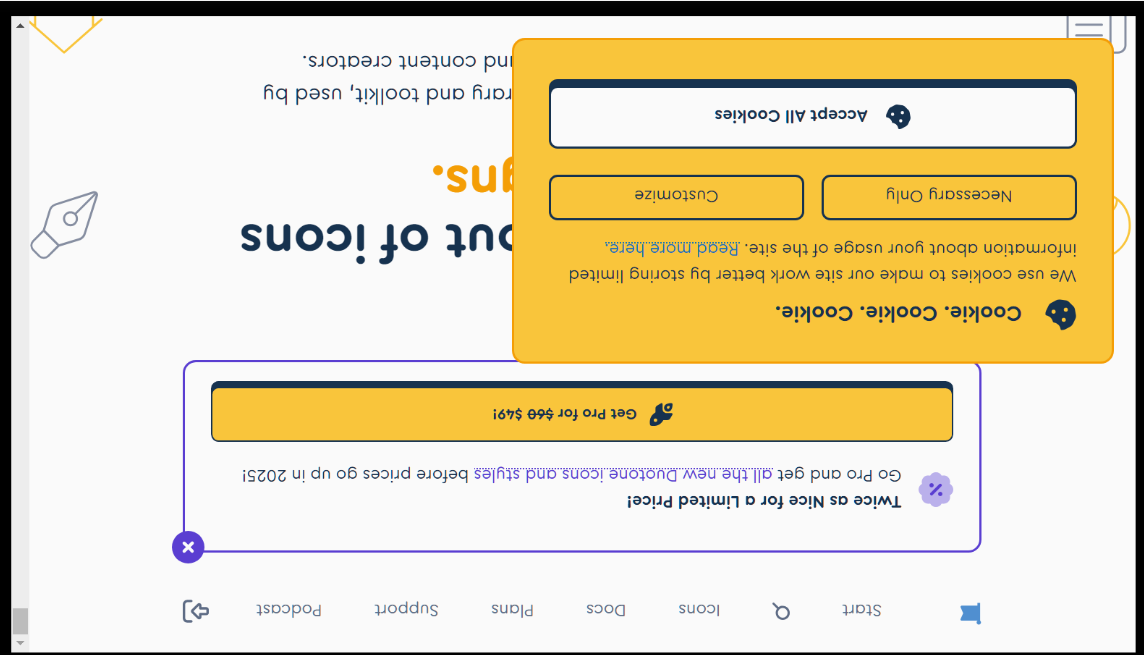}
\caption{\centering Perturbed Input Image for \textit{fontawesome.com}}
\label{fig:fontawesome_noise}
\end{figure}

\newtext{
\section{Perturbation of DeepRare Input Images}}

\label{app:noise}
\newtext{To get more robust estimates of button salience, we inject a small amount of noise into the cookie banner images before inputting them to DeepRare. By randomly injecting noise 32 times (chosen because we have five injections of noise, each with a 50\% chance of selection for each perturbed image) and then averaging button salience scores across the 32 perturbed images, we reduce the chance of DeepRare noise affecting the results. }

\newtext{The five perturbations we consider for each image are 1) shifting the image along the x and y axes, 2) flipping the image over the horizontal axis, 3) flipping the image over the vertical axis, 4) changing the image's brightness, contrast, hue, and saturation with TorchVision's ColorJitter function~\footnote{\url{https://docs.pytorch.org/vision/main/generated/torchvision.transforms.ColorJitter.html}}, and 5) adding Gaussian blur. Each transformation has a 50\% chance of being applied in each perturbed image. For the shift along the x and y axes, a randomly generated shift of up to 15 pixels is chosen in either direction for each axis. For the change in image brightness, contrast, hue and saturation, each is randomly perturbed by $\pm$2\%. We choose 2\% because it is the JND determined for brightness/contrast based on Weber's Law~\cite{pelli2013}. Finally, Gaussian blur is added using the TorchVision function GaussianBlur~\footnote{\url{https://docs.pytorch.org/vision/stable/generated/torchvision.transforms.GaussianBlur.html}}. We aim to add a minimal amount of noise, so we select a kernel size of 3 and the standard deviation for the Gaussian kernel is no greater than 0.02. The generated Gaussian kernel is convolved with the image input to create a slight blurring effect. As an example of these transformations, we display an original input image in Figure~\ref{fig:fontawesome_orig} and a perturbed image in Figure~\ref{fig:fontawesome_noise}.}

\newtext{\section{Salient Object Ranking Methods}}
\label{app:sodrank}
\begin{figure}[t!]
\centering
\includegraphics[width=0.9\linewidth]{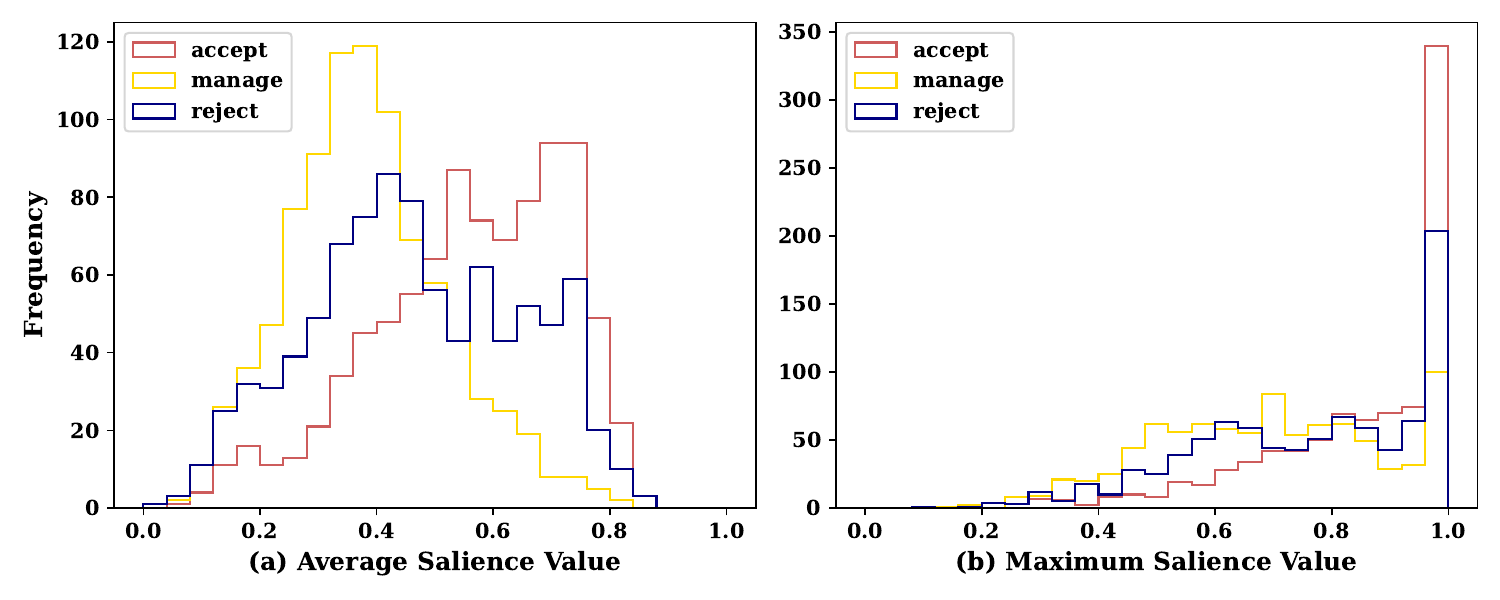}
\caption{\centering Average (a) and Maximum (b) Salience Scores by Button Type}
\label{fig:avgsal}
\end{figure}

\begin{figure}[b!]
\centering
\includegraphics[width=0.82\linewidth]{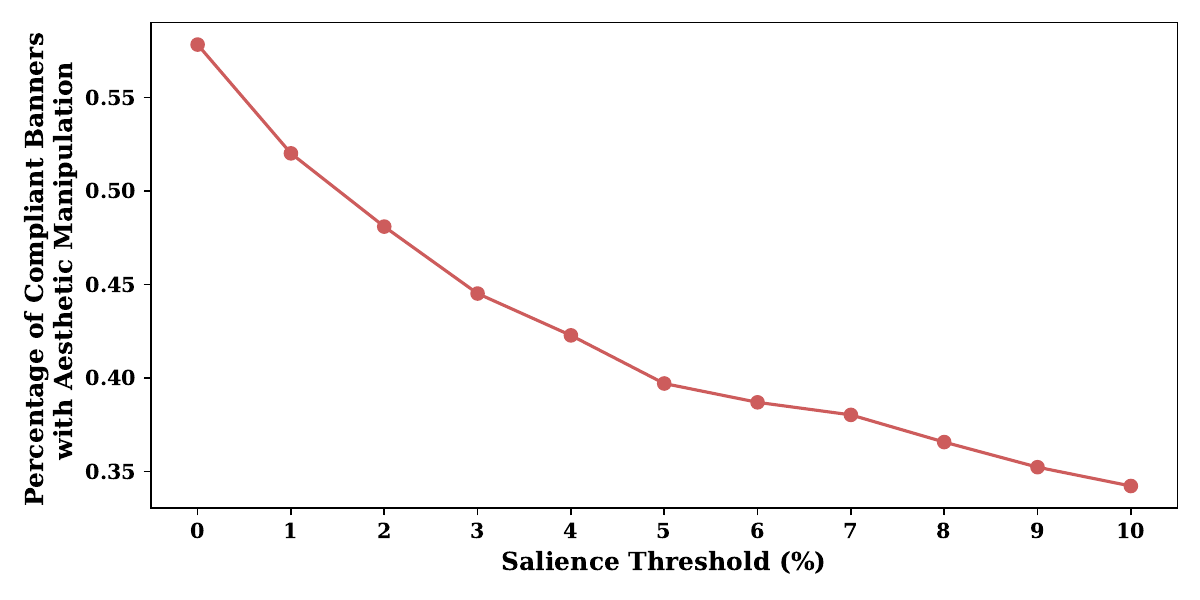}
\caption{\centering Aesthetic Manipulation Prevalence as a Function of Threshold \%}
\label{fig:thresh}
\end{figure}

\begin{figure}[t!]
\centering
\includegraphics[width=0.75\linewidth]{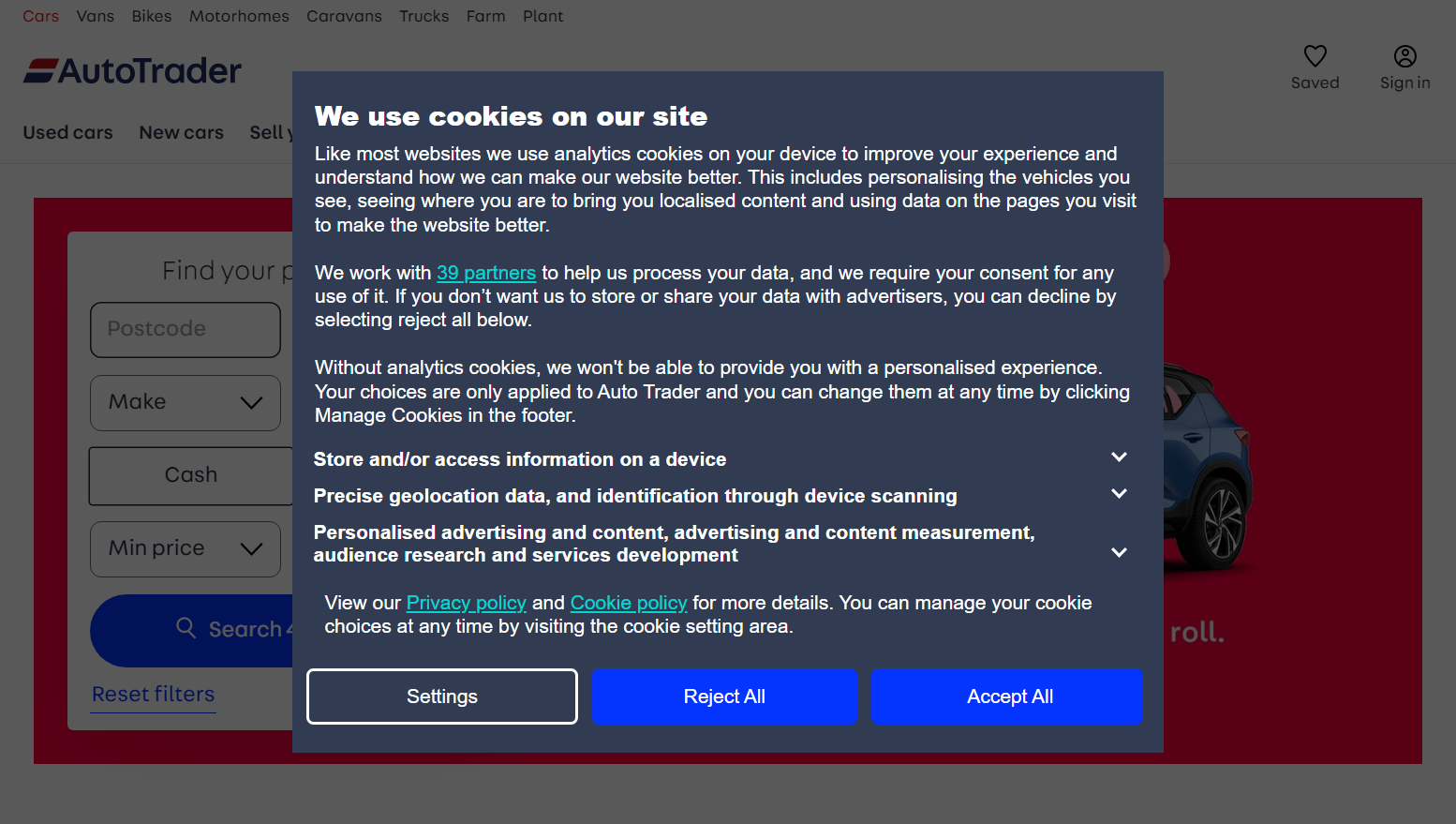}
\caption{\centering False Identification of Aesthetic Manipulation without Threshold}
\label{fig:autotrader}
\end{figure}

\newtext{We experiment with transforming pixel-level salience scores into button-level salience scores by using either the average salience scores or the maximum salience score of all pixels belonging to a button. These methods are based in salient object ranking literature~\cite{kalashsalience,liusalience}. We display the distribution of salience scores for the accept, reject, and manage buttons in compliant banners using each of the two methods in Figure~\ref{fig:avgsal}. Both histograms show that at lower scores, the most frequent button is manage, followed by reject, and then accept. At higher scores, this ordering reverses, and accept buttons are most frequent, with manage buttons being the least frequent. While both methods clearly indicate that across the whole dataset accept buttons are the most salient, followed by reject buttons and then manage buttons, we combine the two scores into one because the two scores capture different elements of salience (as described in Section~\ref{subsec:sod}).}

\section{Threshold Justification}
\label{app:threshold}
\begin{figure}[b!]
\centering
\includegraphics[width=0.75\linewidth]{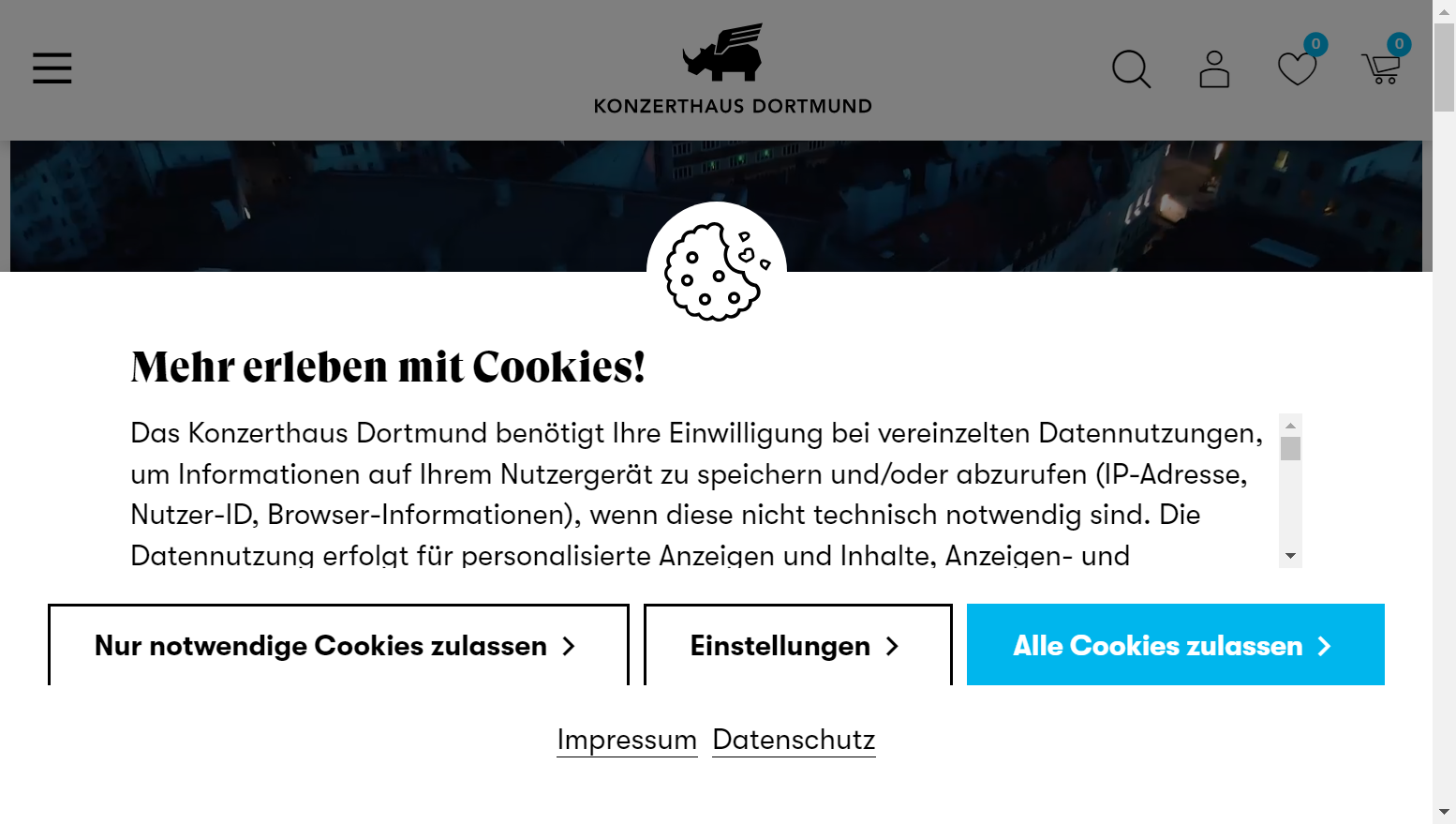}
\caption{\centering Aesthetic Manipulation Not Captured by 10\% Threshold}
\label{fig:konzerthaus}
\end{figure}
\newtext{We choose a 7\% threshold for determining whether one button is more salient than another. The reason for adding any threshold is that we do not want to claim there is aesthetic manipulation when it is not perceived by the user. Although we base our threshold selection in Weber's Law from psychophysics, it is an imperfect, conservative selection based on the threshold for other visual stimuli (e.g., contrast). In Figure~\ref{fig:thresh}, we plot the prevalence of aesthetic manipulation in compliant cookie banners as a function of the chosen threshold. If no threshold is used (i.e., 0\%), we capture banners where the difference in salience is barely noticeable. For example, our method detects a difference of 0.15\% in the accept and reject button salience scores for Figure~\ref{fig:autotrader}, despite the buttons being formatted the same. While the range of JND thresholds for other related visual stimuli (e.g., size, contrast, brightness) stays under 7\%, it is possible that visual salience could have a slightly higher threshold. We report results for thresholds up to 10\%, but caution against defaulting to higher thresholds because it will miss obvious cases of aesthetic manipulation. For example, Figure~\ref{fig:konzerthaus} clearly shows button highlighting (the rightmost button is ``accept'' and leftmost button is ``reject''), but our method only scores the accept button as 8.6\% more salient than the manage and reject buttons. Therefore it would not be recognized by a threshold of 9\% or more. Thus, for our use case of estimating the prevalence of aesthetic manipulation in compliant cookie banners, a 7\% threshold is best supported theoretically and empirically. Future users of this method could utilize more strict (or lax) thresholds depending on their use case. They may also experiment with more specialized detection methods using SOD model scores (e.g., if the difference in either average or maximum salience score exceeds 7\%, label it as aesthetic manipulation). }
\end{document}